\documentclass[letterpaper,titlepage,11pt]{article}
\usepackage{hyperref}
\usepackage{amssymb,amsmath,amsfonts}
\usepackage{epsfig}
\usepackage{graphicx}

\def\clock{{\count0=\time
           \divide\count0 60
           \ifnum\count0<10 0\fi\the\count0
           \multiply\count0 -60 \advance\count0 \time
           :\ifnum\count0<10 0\fi \the\count0
         }}
\newcommand{\timestamp}{{\small\vbox{\hbox{\tt\jobname.tex}
\hbox{\the\day/\the\month/\the\year, \clock}}}}

\newcommand{\be}{\begin{equation}} \newcommand{\ee}{\end{equation}}
\newcommand{\bea}{\begin{eqnarray}} \newcommand{\eea}{\end{eqnarray}}

\def\be{\begin{equation}}
\def\ee{\end{equation}}
\def\beq{\begin{eqnarray}}
\def\eeq{\end{eqnarray}}
\def\ie{$i.e.$ }

\newcommand{\CO}{\mathcal{O}}

\newcommand{\CM}{\mathcal{M}}

\newcommand{\R}{\mathbb{R}}

\newcommand{\lspa}{\ \ ,\ \ \ \ }
\newcommand{\spa}{\,, \ \ }
\newcommand{\rht}{\tilde{\rho}}
\newcommand{\tht}{\tilde{\theta}}

\numberwithin{equation}{section}
\begin{document}

\begin{titlepage}

\rightline{\vbox{\small\hbox{\tt June 2007} }} \vskip 1.6cm

\centerline{\LARGE \bf
Multi-black hole configurations on the cylinder}

\vskip 1.7cm \centerline{\bf \'Oscar J. C. Dias${}^a$, Troels Harmark${}^b$,
Robert C. Myers${}^{c,d,e}$, Niels A. Obers${}^b$} \vskip 0.5cm

\centerline{\sl  $^a$
Departament de F\'{\i}sica Fonamental, Universitat de Barcelona}
\centerline{\sl Av. Diagonal 647,
E-08028 Barcelona, Spain}
\vskip 0.5cm
\centerline{\sl $^b$The Niels Bohr Institute} \centerline{\sl Blegdamsvej 17, 2100
Copenhagen \O, Denmark}
\vskip 0.5cm
\centerline{\sl $^c$ Perimeter Institute for Theoretical Physics}
\centerline{\sl Waterloo, Ontario N2L 2Y5, Canada}
\vskip 0.5cm
\centerline{\sl $^d$  Department of Physics, University of Waterloo}
\centerline{\sl Waterloo, Ontario N2L 3G1, Canada }
\vskip 0.5cm
\centerline{\sl $^e$ Kavli Institute for Theoretical Physics, University of California}
\centerline{\sl Santa Barbara, CA 93106-4030, USA}

\vskip 0.3cm

\centerline{\small\tt odias@ub.edu, harmark@nbi.dk, rmyers@perimeterinstitute.ca, obers@nbi.dk}

\vskip 1.0cm

\centerline{\bf Abstract} \vskip 0.2cm \noindent We construct the
metric of new multi-black hole configurations on a $d$-dimensional
cylinder $\R^{d-1} \times S^1$, in the limit of small total mass (or
equivalently in the limit of a large cylinder). These solutions are
valid to first order in the total mass and describe configurations
with several small black holes located at different points along the
circle direction of the cylinder. We explain that a static
configuration of black holes is required to be in equilibrium such
that the external force on each black hole is zero, and we examine
the resulting conditions. The first-order corrected thermodynamics
of the solutions is obtained and a Newtonian interpretation of it is
given. We then study the consequences of the multi-black hole
configurations for the phase structure of static Kaluza-Klein black
holes and show that our new solutions imply continuous
non-uniqueness in the phase diagram. The new multi-black hole
configurations raise the question of existence of new non-uniform
black strings. Finally, a further analysis of the three-black hole
configuration suggests the possibility of a new class of static
lumpy black holes in Kaluza-Klein space.

\vskip 0.5cm
%\leftline{\timestamp}

\end{titlepage}

\pagestyle{empty}
\small
%\scriptsize
\tableofcontents
\normalsize
%\newpage

\pagestyle{plain}
\setcounter{page}{1}

%%%%%%%%%%%%%%%%%%%%%%%%%%%%%%%%%%%%%%%%%%%%%%%5
\section{Introduction}
%%%%%%%%%%%%%%%%%%%%%%%%%%%%%%%%%%%%%%%%%%%%%%%%

Black holes in four-dimensional General Relativity have a very
simple phase structure. The uniqueness theorems for pure gravity
assert that the only possible stationary black hole solution
 for a given mass and angular momentum is the Kerr black hole.

For higher-dimensional General Relativity, the situation is vastly
different. In particular, if we imagine that we live in a world
which is five dimensional with the extra dimension curled up on a
circle, the relevant black hole solutions are those which asymptote
to four-dimensional Minkowski-space times a circle ($\CM^4\times
S^1$), \ie the five-dimensional Kaluza-Klein space-times. The phase
structure of such black holes has been shown to be very rich
 and contains phases
with event horizons of different topology and even phases where
Kaluza-Klein bubbles are attached to black holes
\cite{Kol:2004ww,Harmark:2005pp,Harmark:2007md}. More generally, we
get a similarly rich phase structure for the case of black holes
asymptoting to $d$-dimensional Minkowski-space times a circle
($\CM^d\times S^1$) with $d \geq 4$.%
\footnote{Note that the case $\CM^3 \times S^1$ studied in
\cite{Myers:1987rx,Bogojevic:1991hv,Korotkin:1994dw,Frolov:2003kd}
is different due to the high amount of symmetry.} The spatial part
of this space-time is a $d$-dimensional cylinder $\R^{d-1} \times
S^1$.

The two static black hole phases which most obviously should appear
for $\CM^d \times S^1$ are the localized black hole phase, which for
small mass behaves as a $d+1$ dimensional Schwarzschild black hole,
and the uniform black string corresponding to a $d$-dimensional
Schwarzschild black hole times a circle. For the uniform string
phase, the metric is known exactly. The most interesting feature of
the uniform string is the Gregory-Laflamme instability
\cite{Gregory:1993vy,Gregory:1994bj} which is a long wave-length
gravitational instability of the solution (see \cite{Harmark:2007md}
for a review). From this instability, it follows that the uniform
string for a certain mass has a marginal mode. From this marginal
mode emanates a new branch of solutions which are non-uniform
strings, \ie solutions with same topology of the event horizons as
the uniform strings but without translational symmetry around the
circle. These new solutions have been studied numerically in
\cite{Gubser:2001ac,Wiseman:2002zc,Harmark:2003dg,Sorkin:2004qq,Kleihaus:2006ee,Sorkin:2006wp}.

For the localized black hole phase, here dubbed the {\sl black hole
on the cylinder} phase, the metric is not known analytically.
However, for small black holes on the cylinder the first order part
of the metric has been found
\cite{Harmark:2002tr,Harmark:2003yz,Gorbonos:2004uc,Gorbonos:2005px}
and also the second order solution has been studied
\cite{Karasik:2004ds,Chu:2006ce}. Finite-size black holes on the
cylinder have instead been studied numerically
\cite{Sorkin:2003ka,Kudoh:2003ki,Kudoh:2004hs}. This study has
revealed the interesting result that the black hole on the cylinder
phase meets the non-uniform string phase in a topology changing
transition point
\cite{Kol:2002xz,Wiseman:2002ti,Kol:2003ja,Sorkin:2006wp}.

In this paper, we find and study new solutions for multi-black hole
configurations on the cylinder. These solutions describe
configurations with several small black holes located at different
points along the circle direction of the cylinder $\R^{d-1} \times
S^1$. The location of each black hole are such that the total force
on each of them is zero, ensuring that they are in equilibrium. It
is moreover necessary for being in equilibrium that the black holes
are all located in the same point in the $\R^{d-1}$ part of the
cylinder.

The metrics that we find are solutions to the Einstein equations to
first order in the mass. More precisely, we work in a regime where the gravitational
interaction between any one of the black holes and the others (and their images on the
circle) is small.
Thus, our solutions describe the small mass
limit of these multi-black hole configurations on the cylinder, or
equivalently they can be said to describe the situation where the
black holes are far apart. The technique used for solving the
Einstein equations is the one developed in \cite{Harmark:2003yz} for
small black holes on the cylinder based on an ansatz for the metric
found in \cite{Harmark:2002tr}.

A subset of the multi-black hole configurations have already been
studied in the literature. These are the so-called copies of the
black hole on the cylinder solutions
\cite{Horowitz:2002dc,Harmark:2003eg,Harmark:2003yz}. This class of
solutions corresponds to the special situation in which a number of
black holes of the same size are spread with equal distance from
each other on the circle.

The existence of these new solutions have striking consequences for
the phase structure of black hole solutions on $\CM^d \times S^1$.
It means that one can for example start from a solution with two
equal size black holes, placed oppositely to each other on the
cylinder, and then continuously deform the solution to be
arbitrarily close to a solution with only one black hole (the other
black hole being arbitrarily small in comparison). Thus, we get a
continuous span of classical static solutions for a given total
mass. This means for static black hole solutions on $\CM^d \times
S^1$ we have in fact a continuous non-uniqueness of solutions.
Continuous non-uniqueness for black holes has also been found when
one attaches Kaluza-Klein bubbles to black holes
\cite{Elvang:2004iz}, and has furthermore been found for other
classes of black hole solutions
\cite{Emparan:2004wy,Elvang:2007rd,Iguchi:2007is,Elvang:2007hg}. In
particular, this has the consequence that if we would live on $\CM^4
\times S^1$ then from a four-dimensional point of view one would
have an infinite non-uniqueness for static black holes of size
similar to the size of the extra dimension, thus severely breaking
the uniqueness of the Schwarzschild black hole.

Another consequence of the new multi-black hole configurations of
this paper is for the connection to uniform and non-uniforms strings
on the cylinder. As mentioned above, there is evidence that the
black hole on the cylinder phase merges with the non-uniform black
string phase in a topology changing transition point. It follows
from this that the copies of black hole on the cylinder solution
merge with the copies of non-uniform black strings. However, with
our new solutions, we add a continuous span of solutions connected
to the copies of the black hole on the cylinder. Therefore, it is
natural to ask whether the new solutions also merge with non-uniform
black string solutions in a topology changing transition point. If
so, it probes the question whether there exist, in addition to
having new black hole on the cylinder solutions, also new
non-uniform black string solutions. Thus, the new solutions of this
paper presents a challenge for the current understanding of the
phase diagram for black holes and strings on the cylinder.

Another connection between strings and black holes on the cylinder
is that a Gregory-Laflamme unstable uniform black string is believed
to decay to a black hole on the cylinder (when the number of
dimensions is less than the critical one \cite{Sorkin:2004qq}).
However, the new solutions of this paper means that one can imagine
them as intermediate steps in the decay.

The solutions presented in this paper are clearly in an unstable
equilibrium. Any small change in the position of one of the black
holes on the cylinder will mean that the black holes will go even
further out of balance, and the endpoint of this instability will
presumably be a single black hole on the cylinder. Nevertheless, one
can argue for their existence for example by imagining two equal
size black holes on the cylinder, and then having mass thrown
towards only one of the black holes in the same way from both sides
of the black hole, $i.e.$ that the solutions keep the inversion
symmetry around both of the black holes. Then the matter will
increase the size of one of the black holes, leaving the other of
the same size.

The construction of multi-black hole solutions also enables us to
examine the possibility of further new types of black hole solutions
in Kaluza-Klein spacetimes. In particular, analysis of the
three-black hole configuration suggests the possibility that new
static configurations may exist that consist of a lumpy black hole
(\ie `peanut-like' shaped black objects), where the non-uniformities are
supported by the gravitational stresses imposed by an external field.

The outline of this paper is as follows. In Section
\ref{sec:construction} we construct the new multi-black hole
configurations on the cylinder to first order in the total mass of
the system. In Section \ref{sec:equilibrium} the equilibrium
condition for these configurations is explored, and a copying
mechanism is presented that generates new equilibrium configurations
from known ones. The first-order corrected thermodynamics of the
multi-black hole solutions is given and analyzed in Section
\ref{sec:thermodynamics}. We then present in Section
\ref{sec:phasediag} the multi-black hole configurations in the phase
diagram for Kaluza-Klein black holes, together with the already
known black hole and black string solutions. Section
\ref{sec:details23bh} contains a more detailed analysis of the two
simplest multi-black hole configurations, namely with two and three
black holes. Finally, Section \ref{sec:conclusions} contains a
summary of our results, a discussion on its implications for
possible new black hole and string phases and open problems. This
concluding section also discusses in the context of an analogue
fluid model a possible, but more speculative, relation of the
multi-black hole configurations to configurations observed in the
time evolution of fluid cylinders. Appendix A contains formulae that
are used to compute thermodynamic quantities for the case of two
unequal mass black holes on a cylinder.

%%%%%%%%%%%%%%%%%%%%%%%%%%%%%%%%%%%
\section{\label{sec:construction}Construction of multi-black hole configurations on the cylinder}
%%%%%%%%%%%%%%%%%%%%%%%%%%%%%%%%%%%

In this section we construct explicitly new solutions describing
multi-black hole configurations on the cylinder, in the limit when
the total mass of the black holes is small.

\subsection{General idea and starting point}

In the following we shall construct new solutions for multi-black hole configurations on the $d$-dimensional cylinder $\R^{d-1} \times S^1$. The solutions are static and they describe configurations with several small black holes located at different points of the cylinder $\R^{d-1} \times S^1$.

We require that all of the black holes are placed in the same point of the $\R^{d-1}$ part of the cylinder. This is necessary in order to have equilibrium. Since all the black holes are placed in the same point of $\R^{d-1}$ we can require the solution to be spherically symmetric on  $\R^{d-1}$. Since the solutions should solve the vacuum Einstein equations, the spherical symmetry has the consequence that we can write the metric for the multi-black hole configuration using the ansatz \cite{Harmark:2002tr,Wiseman:2002ti,Harmark:2003eg}
\begin{eqnarray}
 \label{ansatz}
 ds^2 = - f dt^2 + \frac{A}{f} dR^2 +
\frac{A}{K^{d-2}} dv^2 + K R^2 d\Omega_{d-2}^2 \spa \qquad f = 1 -
\frac{R_0^{d-3}}{R^{d-3}} \ ,
\end{eqnarray}
where $A(R,v)$ and $K(R,v)$ are functions of the two coordinates $R$
and $v$.  As we shall see more explicitly below, the event horizons
for the black holes are all placed at $R=R_0$. For simplicity, we
set the radius of the cylinder to be 1. Thus, the $R$ and $v$
coordinates can be thought of as being measured in units of the
radius of the cylinder. The $v$ coordinate is periodic with period
$2\pi$ \cite{Harmark:2002tr}. For $R \gg 1$, we are in the
asymptotic region where the metric asymptotes to the flat cylinder
metric
\begin{equation}
\label{cylcoord} ds^2 = -dt^2 + dr^2 + r^2 d\Omega_{d-2}^2 + dz^2 \ ,
\end{equation}
where $z$ is periodic with period $2\pi$. Thus, we require that $A(R,v) \rightarrow 1$ and $K(R,v) \rightarrow 1$ for $R \rightarrow \infty$, and we see that $R/r \rightarrow 1$ and $v/z \rightarrow 1$ for $R \rightarrow \infty$.

We construct in the following the metric for multi-black hole
configurations on the cylinder $\R^{d-1} \times S^1$ in the limit
where each of the black holes are small relatively to the distance
between them. To this end, we employ the methods of
\cite{Harmark:2003yz} to find the solution to leading order in the
limit of small total mass. One can equivalently use the methods of
\cite{Gorbonos:2004uc,Gorbonos:2005px} to construct the metric.

We proceed in the following to construct the solution in three steps:
\begin{itemize}
\item Step 1: We find a metric corresponding to the Newtonian gravitational potential sourced by a configuration of small black holes on the cylinder. This metric is valid in the region $R \gg R_0$.
\item Step 2: We consider the Newtonian solution close to the sources, \ie in
the overlap region $R_0 \ll R \ll 1$.
\item Step 3: We find a general solution near a given event horizon and match this solution to the metric in the overlap region found in Step 2. The resulting solution is valid in the region $R_0 \leq R \ll 1$.
\end{itemize}
With all these three steps implemented, we have a complete solution
for all of the spacetime outside the event horizon.

Note that the solutions that we find below generalize the previously studied case of a single
 black hole on a $d$-dimensional cylinder
 \cite{Harmark:2002tr,Harmark:2003yz,Gorbonos:2004uc,Gorbonos:2005px}, \ie a black hole
 with $S^{d-1}$ topology in a $d+1$ dimensional Kaluza-Klein space-time
$\CM^d \times S^1$, $\CM^d$ being $d$-dimensional Minkowski space.
The solutions furthermore generalize the so-called copies of the single-black hole on the
 cylinder solution, corresponding to copying the solution several times across the cylinder,
 thus giving a multi-black hole solution where each of the black holes have the same mass
 and with the black holes placed equidistantly along the circle direction of the cylinder
 \cite{Horowitz:2002dc,Harmark:2003eg}.

\subsection{Step 1: The Newtonian region}
\label{sec:step1}

We construct here the linearized solution for the multi-black hole
configuration in the region $R \gg R_0$ away from the event
horizons. We require the black holes to be small such that they
interact through Special Relativistic gravity (\ie a
Lorentz-invariant extension of Newtonian gravity). In such a Special
Relativistic gravity theory we have a potential for each component
of the energy-momentum tensor that we turn on. For static solutions
on the cylinder it is well-known that the two relevant components of
the energy-momentum tensor are the mass density $\varrho = T_{00}$
and the binding energy (tension) $b = - T_{zz}$
\cite{Harmark:2003dg}. These components source the two gravitational
potentials
\begin{equation}
\label{pots} \nabla^2 \Phi = 8\pi G_{\rm N} \frac{d-2}{d-1}
\varrho \spa \qquad \nabla^2 B = - \frac{8\pi G_{\rm N}}{d-1} b \ ,
\end{equation}
where $G_{\rm N}$ is the $(d+1)$-dimensional Newton constant. From the components of the energy-momentum tensor one finds the total mass $M$ and the relative binding energy (also known as the relative tension) $n$ as \cite{Harmark:2003dg}
\begin{equation}
M = \int d^d x \, \varrho (x) \spa \qquad n =  \frac{1}{M} \int d^d
x \, b(x) \ .
\end{equation}
In the limit of small total mass, we have that the relative binding
energy goes to zero for a single black hole, \ie $n\rightarrow 0$
for $M \rightarrow 0$ \cite{Harmark:2003yz}. From this we have that
$B/(G_{\rm N}M) \rightarrow 0$ for $M \rightarrow 0$. Since $\Phi$
is proportional to $G_{\rm N}M$, this means that we can neglect the
binding energy potential $B$ as compared to the mass density
potential $\Phi$, since $B$ goes like $(G_{\rm N} M)^2$ for small
masses. With this, we see that we only need to consider the
potential $\Phi$, and we thus see that we are considering Newtonian
gravity, with the only potential being the potential $\Phi$ sourced
by the mass density.

We now proceed to find the Newtonian gravity potential $\Phi$. We
consider a configuration of $k$ black holes placed on the cylinder.
We write $M$ as the total mass of all of the black holes. Define
$\nu_i$ as the fraction of mass of the $i^{\rm th}$ black hole, \ie
\begin{equation}
\label{mu} M_i = \nu_i M \,,\qquad \sum_{i=1}^k \nu_i = 1 \,,
\end{equation}
where $M_i$ is the mass of the $i^{\rm th}$ black hole. Note that $0
< \nu_i \leq 1$. As discussed above, we place the black holes in the
same point of the $\R^{d-1}$ part of the cylinder. This corresponds
to $r=0$ in the $(r,z)$ coordinates of the cylinder
\eqref{cylcoord}. Let now $z_i^*$ be the $z$ coordinate for the
$i^{\rm th}$ black hole with mass $\nu_i M$. We can then solve the
equation for $\Phi$ in \eqref{pots} as
\begin{eqnarray}
\Phi (r,z) = - \frac{8 \pi G_N M  }{(d-1) \Omega_{d-1}}
F (r,z) \,,
 \label{Phi}
\end{eqnarray}
with
\begin{eqnarray}
F (r,z) = \sum_{i=1}^k \sum_{m=-\infty}^\infty \frac{\nu_i}{ [ r^2 +
(z - z_i^* - 2 \pi m )^2]^{\frac{d-2}{2}}}  \,.
 \label{def F}
\end{eqnarray}
The potential \eqref{Phi} thus describes the Newtonian gravitational
potential sourced by our multi-black hole configuration. One can
also write the function $F(r,z)$ as the Fourier series
\begin{eqnarray}
F (r,z)=\frac{k_d}{r^{d-3}} {\biggl (} 1+2 \sum_{i=1}^k \nu_i
\sum_{m=1}^{\infty} h(m r)\cos[m(z-z_i^*)] {\biggr )} \ .
 \label{fourier F}
\end{eqnarray}
Here the constant $k_d$ is defined as
\begin{eqnarray}
\label{defkd}
k_d \equiv \frac{1}{2\pi} \frac{d-2}{d-3}
\frac{\Omega_{d-1}}{\Omega_{d-2}} \, ,
\end{eqnarray}
and $h(x)$ as
\begin{equation}
\label{defh} h(x) = 2^{-\frac{d-5}{2}} \frac{1}{\Gamma \left(
\frac{d-3}{2} \right)} x^{\frac{d-3}{2}} K_{\frac{d-3}{2}}(x) \ ,
\end{equation}
where $h(0)=1$, and $K_s(x)$ is the
modified Bessel function of the second kind (in standard notation \cite{abram}).
 For $r \rightarrow \infty$ we see that
\begin{equation}
F(r,z) \simeq \frac{k_d}{r^{d-3}} \ .
\end{equation}
Inserting this in \eqref{Phi} we verify that the potential $\Phi$ has the correct asymptotic behavior for $r\rightarrow \infty$ of a Newtonian potential on the cylinder describing an object with total mass $M$.

We now proceed to find a metric in the form of the ansatz
\eqref{ansatz} describing the linearized solution of the Einstein
equations corresponding to the potential \eqref{Phi}. We first
notice that in the ansatz \eqref{ansatz} we have that $g_{tt} = - 1
+ R_0^{d-3} / R^{d-3}$. However, to leading order in $G_{\rm N} M$
we have that $g_{tt} = - 1 - 2\Phi$. Therefore, we get that
$R^{-d+3}$ is proportional to $\Phi(r,z)$. Demanding furthermore
that $R/r$ for $r \rightarrow \infty$, we are lead to define $R$ as
function of $r$ and $z$ as \cite{Harmark:2002tr}
\begin{eqnarray}
R(r,z) = \left[ \frac{k_d}{F(r,z)} \right]^{\frac{1}{d-3}} \, .
 \label{def R}
\end{eqnarray}
Thus, we see that in order for the linearized metric to fit into the
ansatz  \eqref{ansatz}, we need to define $R$ as \eqref{def R} for
the flat space metric. The choice of $R$ \eqref{def R} is consistent
with having the horizon at $R=R_0$ since we see that defining $R$ in
terms of $F(r,z)$ means that we are defining $R$ to be constant on
the equipotential surfaces of $\Phi$ \cite{Harmark:2002tr}. Since
\eqref{def R} defines $R$ for the flat space metric, we need also to
find a corresponding $v(r,z)$ for the flat space limit of the ansatz
\eqref{ansatz}. One can check, using the flat space metric
\eqref{cylcoord} in cylinder coordinates $r$ and $z$, that in order
to obtain a diagonal metric in the $R$ and $v$ coordinates, we need
$v$ to obey the partial differential equations
\cite{Harmark:2002tr}
\begin{eqnarray}
\partial_r v=\frac{r^{d-2}}{(d-3)k_d}\partial_z F(r,z)\,, \qquad\qquad
\partial_z v=-\frac{r^{d-2}}{(d-3)k_d}\partial_r F(r,z)   \,.
 \label{def v}
\end{eqnarray}
Using the Fourier expansion \eqref{fourier F} of $F(r,z)$ we find
the following explicit solution for $v(r,z)$
\begin{eqnarray}
v = z+ 2 \sum_{i=1}^k  \nu_i  \sum_{m=1}^{+\infty} \sin[m
(z-z_i^*)]\left[\frac{1}{m} h(m r)- \frac{1}{d-3}r h'(m r) \right]
\, ,
 \label{fourier v}
\end{eqnarray}
where $h'(x) \equiv \partial h(x)/\partial x$.
We see that $v/z \rightarrow 1$ as required above.
Given the two coordinates $R$ and $v$ defined in \eqref{def R} and \eqref{fourier v} in terms of $r$ and $z$, we can now find the corresponding flat space metric that can be written in the ansatz \eqref{ansatz}. We find the flat space metric
\begin{eqnarray}
ds^2=-dt^2+ A_0 dR^2+\frac{A_0}{K_0^{d-2}}dv^2+K_0 R^2
d\Omega_{d-2}^2\,,
 \label{metric flat}
\end{eqnarray}
with the function $A_0(r,z)$ and $K_0(r,v)$ given by
\begin{eqnarray}
\hspace{-0.5cm} A_0(r,z)=(d-3)^2 k_d^{-\frac{2}{d-3}}
\frac{F(r,z)^{2\frac{d-2}{d-3}}}{(\partial_r F)^2+(\partial_z F)^2}
\,, \qquad
 K_0(r,z)=r^2 k_d^{-\frac{2}{d-3}}F(r,z)^{\frac{2}{d-3}}
\,.
 \label{K0 A0}
\end{eqnarray}
Using now \eqref{K0 A0} together with \eqref{def R} and \eqref{fourier v}, we can find the two functions $A_0(R,v)$ and $K_0(R,v)$ and we have thereby specified completely the flat space metric \eqref{metric flat}.

With the flat space metric \eqref{metric flat}, as found above from
requiring $g_{tt}$ in the ansatz \eqref{ansatz} to be consistent
with the Newtonian potential \eqref{Phi}, we are now ready to find
the complete metric to first order in $G_{\rm N} M$ in the Newtonian
regime $R \gg R_0$. This problem is solved in general in
\cite{Harmark:2003yz}, and we refer to section 4.1 in that paper for
the details. The upshot is that given the flat space metric
\eqref{metric flat} defined from the Newtonian potential $\Phi$ in
\eqref{Phi}, we can find the correction to first order in $G_{\rm N}
M$ of the functions $A(R,v)$ and $K(r,v)$ as
\begin{eqnarray}
& & A=\left( 1-\frac{1}{(d-2)(d-3)}
 \frac{R_0^{d-3}}{R^{d-3}} \right) A_0
 - \frac{R}{2(d-3)}\frac{R_0^{d-3}}{R^{d-3}}
 \partial_R A_0  \,, \nonumber \\
& & K=\left( 1-\frac{1}{(d-2)(d-3)}
 \frac{R_0^{d-3}}{R^{d-3}} \right) K_0
 - \frac{R}{2(d-3)}\frac{R_0^{d-3}}{R^{d-3}}
 \partial_R K_0   \,.
 \label{A,K-A0,K0}
\end{eqnarray}
Thus, given $A_0(R,v)$ and $K_0(R,v)$, as found above in \eqref{K0
A0}, \eqref{def R} and \eqref{fourier v}, we can find $A(R,v)$ and
$K(R,v)$ to first order in $G_{\rm N} M$, or, equivalently, to first
order in $R_0^{d-3}$. Combining this with the ansatz for the metric
\eqref{ansatz}, we have actually found the metric up to first order
in $R_0^{d-3}$ (\ie in $M$) in the Newtonian region $R \gg R_0$, for
any given distribution of $k$ small black holes on the cylinder.

\subsection{Step 2: The overlap region}
\label{sec:step2}

In the previous Section \ref{sec:step1} we found the metric for any
given distribution of $k$ small black holes on the cylinder to first
order in the total mass. This metric is valid for $R \gg R_0$, \ie
away from the horizon. In this section we examine now this solution
in the region $R_0 \ll R \ll 1$, which we dub the overlap region,
since this is the region where both the Newtonian regime and the
near-horizon solutions are valid. As we shall see below, the
analysis of the solution in the overlap region gives in turn a
restriction on what configurations of black holes that we can find a
metric for, namely that the $k$ black holes should be in equilibrium
with each other with respect to the Newtonian gravitational forces
between them.

Before turning to the first-order corrected metric found in Section
\ref{sec:step1}, we first consider how the potential $\Phi$ looks
when going near the sources, and subsequently how the flat space
metric \eqref{metric flat} behaves. In terms of the flat space
coordinates $R$ and $v$ found in \eqref{def R} and \eqref{fourier
v}, this corresponds to having $R \ll 1$. Note that since we have
$k$ small black holes we have to specify to which of these we are close.
In line with this, it is useful to define for
the $i^{\rm th}$ black hole the spherical coordinates $\rho$ and
$\theta$ by
\begin{equation}
\label{rhotheta} r = \rho \sin \theta \spa z-z^*_i = \rho \cos
\theta \,.
\end{equation}
Notice here that the angle $\theta$ is defined in the interval
$[0,\pi]$. We then conclude from \eqref{rhotheta} that going near
the $i^{\rm th}$ black hole corresponds to having $\rho \ll 1$. We
begin by examining the function $F(r,z)$ in \eqref{def F} near the
$i^{\rm th}$ black hole. In terms of the spherical coordinates
\eqref{rhotheta} we find that
\begin{eqnarray}
F (\rho,\theta) = \nu_i \rho^{-(d-2)} + \Lambda^{(i)} +
\Lambda_1^{(i)} \cos\theta \,\rho + \Lambda_2^{(i)} (d \cos^2\theta
-1) \rho^2 + {\cal O}\left(\rho^3 \right) \,,
 \label{lim F}
\end{eqnarray}
for $\rho \ll 1$, where
\begin{eqnarray}
& & \hskip -.9cm \Lambda^{(i)} =  \nu_i \,\frac{2\zeta (d-2)}{ (2 \pi)^{d-2}}
  +\sum_{\substack{j=1 \\j\neq i } }^k \left\{ \frac{\nu_j}{z_{ij}^{d-2}}+\frac{\nu_j} { (2 \pi)^{d-2}}
  \left [ \zeta \left(d-2,1-\frac{z_{ij}}{2\pi}\right)+\zeta \left(d-2,1+\frac{z_{ij}}{2\pi}\right) \right
  ]\right\} \!\!,\label{lambda} \\
& & \hskip -.9cm  \Lambda_1^{(i)} = (d-2)\sum_{\substack{j=1 \\j\neq i } }^k
\left\{ \frac{\nu_j}{z_{ij}^{d-1}} -\frac{\nu_j}{(2 \pi)^{d-1}}
\left [ \zeta \left(d-1,1-\frac{z_{ij}}{2\pi}\right) - \zeta
\left(d-1,1+\frac{z_{ij}}{2\pi}\right) \right]\right\}  \,,\label{lambda1} \\
& & \hskip -.9cm  \Lambda_2^{(i)} =  \nu_i \,\frac{(d-2)\zeta (d)}{ (2 \pi)^{d}}
  +\frac{d-2}{2} \sum_{\substack{j=1 \\j\neq i } }^k
   \left\{ \frac{\nu_j}{z_{ij}^{d}}+ \frac{\nu_j}{(2 \pi)^{d}}
\left [ \zeta \left(d,1-\frac{z_{ij}}{2\pi}\right) + \zeta
\left(d,1+\frac{z_{ij}}{2\pi}\right) \right]\right\}  \,.
  \label{lambda2}
\end{eqnarray}
Here
\begin{eqnarray}
\zeta(s,1+a)= \sum_{m=1}^{\infty}(m+a)^{-s}\,, \qquad m+a\neq 0\,,
\label{Gzeta}
\end{eqnarray}
is the Generalized Riemann Zeta function and $z_{ij}$ labels the
distance in the $z$ direction between the $j^{\rm th}$ and $i^{\rm
th}$ black hole as follows
\begin{eqnarray}
& & z_{ij}=z_j^*-z_i^* \,, \qquad {\rm if}\:\:\:  0 \leq z_j^*-z_i^*
< 2\pi \,,
\nonumber \\
& & z_{ij}=2\pi+z_j^*-z_i^* \,, \qquad {\rm if}\:\:\: -2\pi\leq
z_j^*-z_i^*<0 \,.\label{defZi}
\end{eqnarray}
We see that this definition ensures that $0 \leq z_{ij} < 2\pi$.

Using now \eqref{lim F}-\eqref{defZi} with \eqref{Phi} one obtains
the behavior of the Newtonian potential $\Phi$ near the $i^{\rm th}$
black hole.

From the potential $\Phi$ for $\rho \ll 1$ obtained by inserting
\eqref{lim F} in \eqref{Phi} we see that  the first term in
\eqref{lim F} corresponds to the flat space gravitational potential due to the
$i^{\rm th}$ mass $M_i = \nu_i M$ and the second term is a constant potential
due to its images and the presence of the other masses and their images.%
\footnote{In particular, the origin of the three terms contributing to
$\Lambda^{(i)}$ in \eqref{lambda} is as follows. The first term comes from the images
of the $i^{\rm th}$ black hole, the second term from the other $k-1$ black holes and
the third term from the images of these.}
Furthermore the third term in \eqref{lim F} is proportional to $\rho \cos \theta = z
- z_i^*$ and therefore this term  gives a
non-zero constant term in $\partial_z \Phi$ if we have that
$\Lambda_1^{(i)}$ given in \eqref{lambda1} is non-zero. This
therefore corresponds to the external force on the $i^{\rm th}$
black hole, due to the other $k-1$ black holes. In Section
\ref{sec:equilibrium} we verify this interpretation.

Since $\Lambda_1^{(i)}$ is proportional to the external force on the
$i^{\rm th}$ black hole, it is clear that one cannot expect a static
solution to exists if $\Lambda_1^{(i)}$ is non-zero, since then the
$i^{\rm th}$ black hole would accelerate along the $z$ axis.
Therefore, the only hope of getting a static solution is if
$\Lambda_1^{(i)}=0$ for all $i=1,2,...,k$, \ie that the external
forces on each of the $k$ black holes are zero. When constructing
our solution, we therefore assume that $\Lambda_1^{(i)}=0$ for all
$i$. From \eqref{lambda1}, we see that this gives conditions on the
relation between the positions $z_i^*$ and the mass ratios $\nu_i$.
We explore these conditions further in Section
\ref{sec:equilibrium}. Note that the equilibrium established with
$\Lambda_1^{(i)}=0$ for all $i$ is an unstable equilibrium, \ie generic small
disturbances in the position of one of the black holes will
disturb the balance of the configuration and result in
the merger of all of the black holes into a single black hole.

We consider now how the flat space metric \eqref{metric flat} looks near the black holes. To this end, it is useful to consider the flat space coordinates $R$ and $v$
found in \eqref{def R} and \eqref{fourier v} near the $i^{\rm th}$
black hole. Using \eqref{lim F}, we see that
\begin{eqnarray}
 \label{lim Rv}
R^{d-3} \simeq \nu_i^{-1} k_d \rho^{d-2} \,, \qquad v\simeq
p_i-\nu_i \,\frac{d-2}{d-3} k_d^{-1} \int_{x=0}^{\theta}dx \, (\sin
x)^{d-2} \,,
\end{eqnarray}
for $\rho \ll 1$, with the number $p_i$ defined as
\begin{eqnarray}
p_i=\pi\,, \quad {\rm for}\:\: i=1\,; \qquad p_i=\pi-2\pi
\sum_{j=1}^{i-1}\nu_{j}\,, \quad {\rm for}\:\: i=2,\cdots,k \,.
 \label{pi}
\end{eqnarray}
Note that $\theta=0$ corresponds to $v=p_i$ and $\theta=\pi$
corresponds to $v=p_i-2\pi\nu_i$. So the range of the coordinate $v$
can belong to one of the $k$ intervals $I_i$ defined as
\begin{eqnarray}
I_i=[p_i-2\pi \nu_i, p_i] \,, \qquad {\rm with } \:\:\:
\bigcup_{i=1}^{k} I_i=[-\pi,\pi] \,,
 \label{range v}
\end{eqnarray}
where the last condition follows from the fact that $\sum_{i=1}^k
\nu_i = 1$. The physical meaning of the intervals \eqref{range v} is
that each of the interval corresponds to one of the black holes. So,
being close to the $i^{\rm th}$ black hole in $(R,v)$ coordinates
corresponds to having $R \ll 1$ and $v \in I_i$. This feature
continues to hold also in the first-order corrected metric.

In order to match the metric in the overlap region to the metric
near the horizons of the black holes, it is natural to change the
ansatz \eqref{ansatz} into a form which resembles more the spherical
coordinates $(\rho,\theta)$, instead of the cylindrical coordinates
$(r,z)$. Given a solution in the form of the ansatz \eqref{ansatz}
with the functions $A(R,v)$ and $K(R,v)$, we define therefore,
relative to the $i^{\rm th}$ black hole, the new coordinates $\rht$
and $\tht$ by \cite{Harmark:2003yz}\footnote{The factor $\nu_i$  in
the second expression of \eqref{defrtht} guarantees that
$\tilde{\theta}=0\leftrightarrow v=p_i$ while $\tilde{\theta}=\pi
\leftrightarrow v=p_i-2\pi\nu_i$. We choose to include the $\nu_i$
in the first line of \eqref{defrtht} to have
$\tilde{\rho}/\rho\rightarrow 1$ when $R\rightarrow 0$; see
\eqref{exptht}.}
\begin{eqnarray}
R^{d-3}= \nu_i^{-1} k_d \tilde{\rho}^{d-2} \spa v=p_i-\nu_i
\,\frac{d-2}{d-3} k_d^{-1} \int_{x=0}^{\tilde{\theta}}dx \, (\sin
x)^{d-2} \,.
 \label{defrtht}
\end{eqnarray}
where $p_i$ is defined in \eqref{pi}, and  $\tilde{\theta}=0$
corresponds to $v=p_i$ while $\tilde{\theta}=\pi$ corresponds to
$v=p_i-2\pi\nu_i$. The coordinates $(\rht,\tht)$ are defined such
that $\rht=\rht(R)$ and $\tht=\tht(v)$ and such that for the flat
space metric we have $\rht \simeq \rho$ and $\tht \simeq \theta$ for
$\rho \ll 1$, as one can see from \eqref{lim Rv}. We define
furthermore the two functions $\tilde{A}( \rht,\tht)$ and
$\tilde{K}(\rht,\tht)$ by
\begin{eqnarray}
A=\frac{(d-3)^2}{(d-2)^2} \left(\nu_i^{-1} k_d \tilde{\rho}
\right)^{-\frac{2}{d-3}}\tilde{A}\,, \qquad
K=\sin^2\tilde{\theta}\left(\nu_i^{-1} k_d \tilde{\rho}
\right)^{-\frac{2}{d-3}}\tilde{K}   \,,
 \label{def A,K tilde}
\end{eqnarray}
and the parameter $\rho_0$ by
\begin{eqnarray}
\rho_0^{d-2}= k_d^{-1} R_0^{d-3}\ ,
 \label{defrho0}
\end{eqnarray}
such that we can write the ansatz \eqref{ansatz} in the alternative
form
\begin{eqnarray}
ds^2=-f dt^2+ \frac{\tilde{A}}{f}d\tilde{\rho}^2 +
\frac{\tilde{A}}{\tilde{K}^{d-2}}\tilde{\rho}^2d\tilde{\theta}^2
 +  \tilde{K}
\tilde{\rho}^2 \sin^2\tilde{\theta} d\Omega_{d-2}^2\,, \qquad
f=1-\frac{\nu_i \rho_0^{d-2}}{\tilde{\rho}^{d-2}} \ .
 \label{newansatz}
\end{eqnarray}
Note that the event horizon for the $i^{\rm th}$ black hole is
located at $\rht=\nu_i^{\frac{1}{d-2}}\rho_0$.

Turning to the flat space metric, corresponding to the zero total
mass limit of the metric for the multi-black hole configuration, we
can reformulate the above results for the $(R,v)$ coordinates in
terms of the $(\rht,\tht)$ coordinates. We write the flat space
limit of the ansatz \eqref{newansatz} as
\begin{eqnarray}
\label{newflat}
 ds^2 = - dt^2 +  \tilde{A}_0 d\rht^2 +
  \frac{\tilde{A}_0}{\tilde{K}_0^{d-2}} \rht^2 d\tht^2 +
 \tilde{K}_0 \rht^2 \sin^2 \tht
d\Omega_{d-2}^2 \ .
\end{eqnarray}
The functions $\tilde{A}_0 (\rht,\tht)$ and $\tilde{K}_0
(\rht,\tht)$ defining the flat space metric \eqref{newflat} are most
easily found using the relations
\begin{eqnarray}
\tilde{A}_0 =  \left[ (\partial_\rho \rht)^2 + \rht^2
\tilde{K}_0^{-(d-2)} (\partial_\rho \tht)^2 \right]^{-1} \,, \qquad
\tilde{K}_0 =   \frac{\rho^2 \sin^2 \theta}{\rht^2 \sin^2 \tht} \ .
\label{tilA0K0aux}
\end{eqnarray}
Implementing now the definitions \eqref{defrtht} and the results
\eqref{lim Rv}, we see that for $\rht \ll 1$ (which is equivalent to
$\rho \ll 1$) we get the expansion%
\footnote{We included here for completeness the $\Lambda_1^{(i)}$
terms although we set $\Lambda_1^{(i)}=0$ in the actual solutions in
order to have a static solution, as discussed above.}
\begin{eqnarray}
& & \rho = \rht \left[ 1 + \frac{\nu_i^{-1} \Lambda^{(i)}}{d-2}\,
\rht^{d-2} + \frac{\nu_i^{-1} \Lambda_1^{(i)} }{d-2} \,\cos\tht \,
\rht^{d-1}  + \CO ( \rht^{d} ) \right] \,,  \nonumber \\
 & & \sin^2 \theta = \sin^2 \tht \left[ 1
 + \frac{2 \nu_i^{-1} \Lambda_1^{(i)} }{(d-1)(d-2)}\, \cos\tht \, \rht^{d-1}
   + \CO (\rht^{d} ) \right] \ . \label{exptht}
\end{eqnarray}
Using this with \eqref{tilA0K0aux}, we find the following expansions
for $\tilde{A}_0 (\rht,\tht)$ and $\tilde{K}_0 (\rht,\tht)$
\begin{eqnarray}
\label{tilAK0}  \tilde{A}_0 =  1 + \frac{2(d-1)\nu_i^{-1}
\Lambda^{(i)}}{d-2}\, \rht^{d-2} + \CO (\rht^d)  \spa
  \tilde{K}_0 =  1 + \frac{2\nu_i^{-1}
\Lambda^{(i)}}{d-2} \, \rht^{d-2} + \CO (\rht^d)  \,,
\end{eqnarray}
for $\rht \ll 1$. We included here the corrections up to order
$\rht^{d-2}$. Note that the next corrections come in at order
$\rht^d$ since here and in the following we have set
$\Lambda_1^{(i)}=0$.

Having understood the flat space metric in the ansatz
\eqref{newansatz} near the $i^{\rm th}$ black hole, we are now ready
to collect all the results and write down a first-order corrected
metric near the $i^{\rm th}$ black hole. First, we note that using
the definition \eqref{def A,K tilde} it follows from the general
form \eqref{A,K-A0,K0} for the first-order corrected metric in the
$(R,v)$ coordinates that we obtain the general form for the
first-order corrected metric in the $(\rht,\tht)$ coordinates,
\begin{eqnarray}
\label{AKfirst} \tilde{A} = \tilde{A}_0 - \frac{\rht}{2(d-2)}
\frac{\nu_i \rho_0^{d-2}}{\rht^{d-2}}
\partial_{\rht} \tilde{A}_0
\spa \qquad \tilde{K} = \tilde{K}_0 - \frac{\rht}{2(d-2)}
\frac{\nu_i \rho_0^{d-2}}{\rht^{d-2}}
\partial_{\rht} \tilde{K}_0 \ .
\end{eqnarray}
Given the full flat space functions $\tilde{A}_0 (\rht,\tht)$ and
$\tilde{K}_0 (\rht,\tht)$, the functions $\tilde{A} (\rht,\tht)$ and
$\tilde{K} (\rht,\tht)$ in \eqref{AKfirst} when inserted in the
ansatz \eqref{newansatz} describe the first-order corrected metric
for a configuration of small black holes in the region $\rht \gg
\rho_0$. Using now the $\rht \ll 1$ expansion of $\tilde{A}_0$ and
$\tilde{K}_0$ found in \eqref{tilAK0} we get the following explicit
expansions of the first-order corrected metric for
$\nu_i^{\frac{1}{d-2}} \rho_0 \ll \rht \ll 1$
\begin{eqnarray}
\label{cortAcortK} \hspace{-0.5cm} \tilde{A} \simeq 1+
\frac{(d-1)\nu_i^{-1} \Lambda^{(i)}}{d-2} \left[ 2\rht^{d-2} - \nu_i
\rho_0^{d-2} \right]
 \,, \qquad \tilde{K} \simeq  1+ \frac{\nu_i^{-1} \Lambda^{(i)}}{d-2} \left[
2\rht^{d-2} - \nu_i \rho_0^{d-2} \right].
\end{eqnarray}
Thus,  the functions \eqref{cortAcortK} with the ansatz
\eqref{newansatz} give the metric of the multi-black hole
configuration in the overlap region $\nu_i^{\frac{1}{d-2}} \rho_0
\ll \rht \ll 1$. In Section \ref{sec:step3}, we shall match this
with the metric in the near-horizon region.

\subsubsection*{Regularity of the solution}

We can now address the regularity of the multi-black hole solution
given the above results for the first order correction. We already
argued above that we need the equilibrium condition
$\Lambda_1^{(i)}=0$ to hold for all $i=1,...,k$, since otherwise the
configuration that we are describing cannot be static. However, this
should also follow from demanding regularity of the solution, since
with a non-zero Newtonian force present on the black hole the only
way to keep it static is to introduce a counter-balancing force
supported by a singularity. Therefore, it is important to examine
the regularity of the solution corresponding to \eqref{AKfirst} with
or without the presence of the $\Lambda_1^{(i)}$ terms.

For a metric in the form of the ansatz \eqref{newansatz}, one can
have singularities for $\tht \rightarrow 0,\pi$, since the metric
component along the $(d-2)$-sphere goes to zero there. A necessary
condition to avoid such singularities is that for $\tht \rightarrow
0,\pi$ the $\tht$ part plus the $(d-2)$-sphere part of the metric
\eqref{newansatz} becomes locally like the metric of a
$(d-1)$-sphere $d\tht^2 + \sin^2 \tht d\Omega_{d-2}^2$ since then
$\tht = 0,\pi$ corresponds to the poles of the $(d-1)$-sphere. This
is only the case provided that
\begin{equation}
\label{regcond} \frac{\tilde{A}}{\tilde{K}^{d-1}} \rightarrow 1 \
\mbox{for} \ \tht \rightarrow 0,\pi\,.
\end{equation}
Therefore, we should examine under which conditions the correction
\eqref{AKfirst} obeys Eq.~\eqref{regcond}. First, let us assume that
the flat space functions $\tilde{A}_0$, $\tilde{K}_0$ obey
Eq.~\eqref{regcond}, \ie $\tilde{A}_0/\tilde{K}_0^{d-1} \rightarrow
1$ for $\tht \rightarrow 0,\pi$. From this one can infer that
$\partial_{\rht} \log \tilde{A}_0 - (d-1) \partial_{\rht} \log
\tilde{K}_0 \rightarrow 0$ for $\tht \rightarrow 0,\pi$. Using this,
it is not hard to check that Eq.~\eqref{regcond} is fulfilled with
$\tilde{A}$ and $\tilde{K}$ given by \eqref{AKfirst}. Thus, in order
to fulfil \eqref{regcond} we only need to check that it is fulfilled
for the flat space metric. This is indeed found to be the case, both
for the $\Lambda^{(i)}$ terms and the $\Lambda^{(i)}_1$ terms. Thus,
the metric is regular at the poles $\tht = 0,\pi$ also with the
external force on the $i^{\rm th}$ black hole being present. This is
presumably because we cannot see the irregularity of the solution at
this order since we can neglect the binding energy, which accounts
for the self-interaction of the solution. Thus, we expect
singularities to appear at second order in the total mass for
solutions which do not obey the equilibrium condition
$\Lambda_1^{(i)}=0$.

\subsection{Step 3: The near-horizon region}
\label{sec:step3}

In Section \ref{sec:step1} we found the metric (to first order in
the mass) for a general multi-black hole configuration in the
Newtonian region $R \gg R_0$. We now complete the metric for the
multi-black hole configuration by finding the metric near the
horizon. This is done by matching with the metric in the overlap
region $R_0 \ll R \ll 1$, as found in Section \ref{sec:step2}.

Take the metric \eqref{newansatz} with \eqref{cortAcortK} which
describes the geometry near the $i^{\rm th}$ black hole, \ie in the
overlap region $\nu_i^{\frac{1}{d-2}} \rho_0 \ll \rht \ll 1$. We
notice here the key point that $\tilde{A}$ and $\tilde{K}$ are
independent of $\tht$. This means that we can assume that
$\tilde{A}$ and $\tilde{K}$ are independent of $\tht$ for
$\nu_i^{\frac{1}{d-2}} \rho_0 \leq \rht \ll 1$. The next step is
therefore to find the most general solution of the vacuum Einstein
equations for a metric of the form \eqref{newansatz} with $\tilde{A}
= \tilde{A} ( \rht)$ and $\tilde{K} = \tilde{K} ( \rht)$, \ie
without any $\tht$ dependence. This gives the result
\cite{Harmark:2003yz}
\begin{eqnarray}
\label{AKw}   \tilde{A}^{- \frac{d-2}{2(d-1)}} =  \tilde{K}^{-
\frac{d-2}{2}} = \frac{1-w^2}{w} \frac{\rht^{d-2}}{\nu_i
\rho_0^{d-2}} + w \ ,
\end{eqnarray}
where $w$ is an arbitrary constant. Note that, setting $w=1$, the
ansatz \eqref{newansatz} with \eqref{AKw} describes the
($d+1$)-dimensional Schwarzschild black hole solution.

We can now fix this constant $w$ by matching the functions
\eqref{AKw} to the behavior of $\tilde{A}$ and $\tilde{K}$ in the
overlap region \eqref{cortAcortK}. This yields
\begin{eqnarray}
\label{thew} w = 1 + \frac{ \Lambda^{(i)}}{2}  \rho_0^{d-2} + \CO
(\rho_0^{2(d-2)} ) \ .
\end{eqnarray}
Thus, using \eqref{AKw} with \eqref{thew} in the ansatz
\eqref{newansatz}, we have obtained the metric for a general
multi-black hole configuration, in the limit of small total mass, in
the near-horizon region $\nu_i^{\frac{1}{d-2}} \rho_0 \leq \rht \ll
1$. Supplementing this with the metric in the Newtonian region $R
\gg R_0$ found in Section \ref{sec:step1}, we see that we have
obtained the full metric for the general multi-black hole
configuration to first order in the mass in the limit of small total
mass.

Inserting \eqref{AKw} and \eqref{thew} in the ansatz
\eqref{newansatz}, we can write the near-horizon metric near the
$i^{\rm th}$ black holes located at $(r,z)=(0,z_i^*)$ as
\begin{eqnarray}
\label{met1} ds^2 = - f dt^2 + f^{-1} G^{-\frac{2(d-1)}{d-2}}
d\rht^2 + G^{-\frac{2}{d-2}} \rht^2 \left( d\tht^2 + \sin^2 \tht \,
d\Omega_{d-2}^2 \right)
 \ ,
\end{eqnarray}
where (up to first order in $\rho_0^{d-2}$)
\begin{eqnarray}
\label{met2} f = 1 - \frac{\nu_i \rho_0^{d-2}}{\rht^{d-2}} \spa
G(\rht) = \frac{1-w^2}{w} \frac{\rht^{d-2}}{\nu_i \rho_0^{d-2}} + w
\spa  w = 1 + \frac{ \Lambda^{(i)} }{2} \rho_0^{d-2}+ \CO (
\rho_0^{2(d-2)} ) \ .
\end{eqnarray}
The horizon is located at $\rht=\nu_i^{\frac{1}{d-2}}\rho_0$ and the
range of $\tht$ is from $0$ to $\pi$.

%%%%%%%%%%%%%%%%%%%%%%%%%%%%%%%%%%%%%%%%%%%%%%%%%%%%%%%%%%%%%%
\section{\label{sec:equilibrium}Equilibrium configurations}
%%%%%%%%%%%%%%%%%%%%%%%%%%%%%%%%%%%%%%%%%%

From the results of Section \ref{sec:construction} we have that near
the $i^{\rm th}$ black hole the gradient of
the gravitational potential along the $z$-direction  is
\begin{equation}
\label{gradphi}
\partial_z \Phi = \frac{8\pi G_{\rm N} M}{(d-1)\Omega_{d-1}} \left(
(d-2) \frac{z-z_i^*}{\rho^d} - \Lambda_1^{(i)} + \CO( \rho ) \right)
\ ,
\end{equation}
for $\rho \ll 1$. The first term is evidently the gravitational
attraction due to the mass of the $i^{\rm th}$ black hole, while the
second term is a net force on the $i^{\rm th}$ black hole, which
originates from the other $k-1$ black holes and their images in the configuration.%
\footnote{The images of the $i^{\rm  th}$ black hole only contribute in
Eq.~\eqref{gradphi} in the terms of $\CO( \rho )$.}
Having such a force on the $i^{\rm th}$ black hole is clearly not
consistent with having a static solution. Therefore, as already
discussed in Section \ref{sec:step2}, we require that the solutions
fulfil the equilibrium condition
\begin{eqnarray}
 \Lambda_1^{(i)} = 0 \, \qquad {\rm for} \:\:\: i=1,...,k  \,.
  \label{EquilCond}
\end{eqnarray}
In Section \ref{sec:EquilSolutions} we explore this condition
further, and we describe a method of how to find configurations, \ie
a set of masses $\nu_i$ and positions $z_i^*$, such that the
equilibrium condition \eqref{EquilCond} is fulfilled. We furthermore
describe in Section \ref{sec:copying} how to generate new
equilibrium configurations from known ones by copying.

As already discussed in Section \ref{sec:step2}, the equilibrium of
the $k$ black holes is unstable towards perturbations in the
positions of the black holes. We compare this physical intuition
with the results for the two-black hole solution in Section
\ref{sec:details2BH}.

%%%%%%%%%%%%%%%%%%%%%%%%%%%%%%%%%%%%%%%%%%%%%%%%%%%%%%%%%%%%%%
\subsection{Construction of equilibrium configurations}
 \label{sec:EquilSolutions}
%%%%%%%%%%%%%%%%%%%%%%%%%%%%%%%%%%%%%%%%%%

In the following we describe a construction method that allows one to
find equilibrium configurations fulfilling \eqref{EquilCond}. While doing so we
further clarify the equilibrium conditions.

Condition \eqref{EquilCond} {\it per se} is not in general
sufficient to identify specific parameters of configurations that
are in equilibrium. In the following we describe a procedure from
which we can obtain an equilibrium configuration given a set of
black hole positions (with some restrictions).

We first note that we can write $\Lambda_1^{(i)}$ as a sum of the potential gradients corresponding to the gravitational force due to each of the $k-1$ other black holes on the $i^{\rm th}$ black hole as%
\footnote{Note that the force on the $i^{\rm th}$ black hole is $\Lambda_1^{(i)} 8\pi G_{\rm N} M /((d-1)\Omega_{d-1})$.}
\begin{eqnarray}
\Lambda_1^{(i)} = \sum_{\substack{j=1,j\neq i } }^k  \nu_j V_{i j}
\,,
  \label{LambdaPotent}
\end{eqnarray}
where $V_{ij}$ corresponds to the gravitational field on the $i^{\rm
th}$ black hole from the $j^{\rm th}$ black hole, given by
\begin{eqnarray}
V_{ij} =  (d-2) \left\{ \frac{1}{z_{ij}^{d-1}} -\frac{1}{(2
\pi)^{d-1}} \left [ \zeta \left(d-1,1-\frac{z_{ij}}{2\pi}\right) -
\zeta \left(d-1,1+\frac{z_{ij}}{2\pi}\right) \right] \right\} \,,
  \label{Vij}
\end{eqnarray}
for $j \neq i$. We can now furthermore define $F_{ij} \equiv \nu_i
\nu_j V_{ij}$ as the Newtonian force on the $i^{\rm th}$ mass due to
the $j^{\rm th}$ mass (and its images as seen in the covering space
of the circle). Of course, to obtain the actual Newtonian force we
have to multiply $F_{ij}$ with $8\pi G_{\rm N} M^2
/((d-1)\Omega_{d-1})$. With this, we can write \eqref{LambdaPotent}
as the condition of zero external force on each of the $k$ masses
\begin{equation}
\label{noforce} \sum_{j=1,j \neq i}^k F_{ij} = 0  \ ,
\end{equation}
for $i=1,...,k$. We can now verify an important property, namely
that Newton's law $F_{ij} = - F_{ji}$ is satisfied. Clearly this is
equivalent to $V_{ij} = - V_{ji}$. From \eqref{Vij} and the
definition \eqref{defZi} of $z_{ij}$ for the $i^{\rm th}$ black
hole, we see that $V_{ij} = - V_{ji}$ follows from the following
identify for the Generalized Zeta function \eqref{Gzeta}
\begin{eqnarray}
\hskip -.3cm \left(\frac{2\pi}{2\pi-z}\right)^{s}-  \zeta
\left(s,\frac{z}{2\pi}\right) + \zeta
\left(s,2-\frac{z}{2\pi}\right)
 = - \left(\frac{2\pi}{z}\right)^{s} + \zeta
\left(s,1-\frac{z}{2\pi}\right) - \zeta
\left(s,1+\frac{z}{2\pi}\right).
  \label{propZeta4}
\end{eqnarray}

We now illustrate our procedure of finding equilibrium configurations
by considering the $k=3$ black hole case. The generalization to an
arbitrary number of black holes is easily done. First, consider a
given set of positions of the black holes $(z_1^*,z_2^*,z_3^*)$.
From these positions we get $V_{ij}$ from \eqref{Vij}. We now want
to find $\nu_1$, $\nu_2$ and $\nu_3$ such that we get an equilibrium
configuration. From \eqref{noforce} we see using $F_{ij}=-F_{ji}$
that there are only two independent equations, which we can write as
$\nu_2 V_{12} + \nu_3 V_{13}=0$ and $-\nu_1 V_{12} + \nu_3 V_{23} =
0$. Using now that $\nu_3=1-\nu_1-\nu_2$, we get the following
result for $\nu_1$, $\nu_2$ and $\nu_3$
\begin{eqnarray}
\label{fromnoforce} \nu_1 = \frac{V_{23}}{V_{12}-V_{13}+V_{23}} \spa
\quad \nu_2 = - \frac{V_{13}}{V_{12}-V_{13}+V_{23}} \spa \quad \nu_3
= \frac{V_{12}}{V_{12}-V_{13}+V_{23}} \,.
\end{eqnarray}
Thus, we see that choosing the positions of the three black holes
gives us $V_{ij}$ which again gives us $\nu_1$ and $\nu_2$ from
\eqref{fromnoforce}, implementing the zero force condition
\eqref{noforce}.

However, it is important to note that we need to impose the physical
requirement of having only positive masses, \ie $0 \leq \nu_i \leq
1$ for all $i$. This again gives restrictions on the positions that
one can choose. For $k=3$ one can check that these restrictions are
satisfied under the fairly mild conditions $z_1^* = 0 < z_2^* < \pi
< z_3^* < 2\pi$ and $z_3^*-z_2^*< \pi$.

The above construction method that we described for $k=3$ can be
extended to configurations with any number of black holes subjected
to some constraints on their relative positions. One then solves the
$k-1$ independent zero force conditions from \eqref{noforce} for the
$k-1$ independent mass parameters $\nu_i$. Note that one can infer
from this way of solving the equilibrium condition \eqref{EquilCond}
that in general a $k$ black
hole configuration has $k$ independent parameters, $e.g.$ the rescaled
mass and the $k-1$ positions.%
\footnote{Note that there are special configurations with a high
amount of symmetry where the mass ratios $\nu_i$ are not fixed given
the positions $z_i^*$. An example of this is the two black hole case
with $z_1^*=0$ and $z_2^*=\pi$. However, the number of independent
parameters is always $k$ for a $k$ black hole configuration, $i.e.$
for the two black hole case the two parameters can be taken to be
$\mu$ and $\nu_1$.} Another way to see that we have $k$ independent
parameters for a configuration with $k$ black holes is to note that
by specifying that $z_1^* < z_2^* < ... < z_k^*$ and by giving the
$k$ absolute masses $\nu_i M$ (or alternatively the rescaled total
mass and $k-1$ of the mass parameters $\nu_i$) we can determine an
equilibrium configuration using the analysis above.

%%%%%%%%%%%%%%%%%%%%%%%%%%%%%%%%%%%%%%%%%%%%%%%%%%%%%%%%%%%%%%
\subsection{New equilibrium configurations by copying}
\label{sec:copying}
%%%%%%%%%%%%%%%%%%%%%%%%%%%%%%%%%%%%%%%%%%%%%%%%%%%%%%%%%%%%%%%%%%%

We described above a general method to build equilibrium
configurations. In this section we consider a way to generate new
equilibrium configurations using already known ones. This is done by
copying the configurations a number of times around the circle. This
generalizes the copies of the single-black hole solution
\cite{Horowitz:2002dc,Harmark:2003eg,Harmark:2003yz}.

We imagine a configuration given with $k$ black holes, specified with the positions $z_i^*$ and masses $\nu_i$, $i=1,...,k$. We assume this configuration is in equilibrium, \ie that \eqref{EquilCond} is satisfied. We also assume that the positions are ordered such that $0 \leq z_i^* < z_{i+1}^* < 2\pi $ for $i=1,...,k-1$. Given now an integer $q$, we can copy this configuration $q$ times, to obtain a new equilibrium configuration as follows. We define%
\footnote{Note that here and in the following we put a hat symbol on
all the functions, parameters and quantities that correspond to the
new configuration that we copied $q$ times.}

\begin{equation}
\hat{z}_{i+nk}^* \equiv \frac{1}{q} ( z_i^* + 2\pi n )     \,,
\qquad \hat{\nu}_{i+nk} \equiv \frac{1}{q} \nu_{i}\,,
\end{equation}
for $i=1,...,k$ and $n=0,...,q-1$. Then $\hat{z}_1,...,\hat{z}_{kq}$
and $\hat{\nu}_1,...,\hat{\nu}_{kq}$ defines a new configuration
with $kq$ black holes. In particular we have that $\sum_{a=1}^{kq}
\hat{\nu}_a = 1$ and that $0 \leq \hat{z}_a^* < \hat{z}_{a+1}^* <
2\pi $ for $a=1,...,kq-1$.

We first verify that the new configuration of $kq$ black holes
obeys the equilibrium conditions \eqref{EquilCond}. Note that this
check is needed only for the first $k$ black holes (out of the $kq$
black holes) since the black hole configuration is symmetric under
the transformation $\hat{z}_a^* \rightarrow \hat{z}_{a+k}^*$,
$\hat{\nu}_a \rightarrow \hat{\nu}_{a+k}$ if we furthermore make the
displacement $z \rightarrow z + 2\pi /q$. Consider therefore the
zero force condition on the $i^{\rm th}$ black hole, with
$i=1,...,k$. Using \eqref{LambdaPotent} we can write this as
\begin{equation}
\label{qcopyeq} \sum_{n=1}^{q-1} \hat{\nu}_{i+nk} \hat{V}_{i,i+nk} +
\sum_{n=0}^{q-1} \sum_{j=1,j\neq i}^k \hat{\nu}_{j+nk}
\hat{V}_{i,j+nk} = 0 \,,
\end{equation}
with $\hat{V}_{ab}$ given by \eqref{Vij}. Here we have split up the
contributions such that the first term corresponds to the copies of
the $i^{\rm th}$ black hole, while the second term corresponds to
the other $k-1$ black holes and their copies. Using now that
$\hat{z}_{i,i+nk} = \frac{2\pi n}{q}$ and $\hat{z}_{i,j+nk} =
\frac{z_{ij}}{q} + \frac{2\pi n}{q}$, as one can infer from the
definition \eqref{defZi}, it is straightforward to verify, with the
aid of the definition of the generalized Zeta function
\eqref{Gzeta}, that we have
\begin{equation}
\sum_{n=1}^{q-1} \hat{V}_{i,i+nk} = 0 \,, \qquad \sum_{n=0}^{q-1}
\hat{V}_{i,j+nk} = q^{d-1} V_{ij} \,.
\end{equation}
Using this, we see that it follows from the equilibrium condition
$\sum_{j=1,j\neq i}^k \nu_j V_{ij} = 0$ for the $k$ black hole
configuration that the equilibrium condition \eqref{qcopyeq} is
satisfied for the $kq$ black hole configuration.

It is useful to consider how one can express the metric for the $q$
copied  configuration in terms of the metric for the $k$ black hole
configuration. To this end, we note that one easily sees from
\eqref{def F} that
\begin{equation}
\label{qcopyF}
\hat{F}(r,z) = q^{d-3} F(qr,qz) \,.
\end{equation}
This gives in turn that $\hat{A}_0 (r,z) = A_0(qr,qz)$ and
$\hat{K}_0 (r,z) = K_0(qr,qz)$.  By carefully using these relations,
we infer that $\hat{A}_0 (R,v) = A_0(qR,qv)$ and $\hat{K}_0 (R,v) =
K_0(qR,qv)$. Therefore, we have from \eqref{A,K-A0,K0} that
\begin{equation}
\hat{A} (R,v) = A(qR,qv) \spa\qquad \hat{K} (R,v) = K(qR,qv) \,.
\end{equation}
From this we can read off the metric for the $q$ copied configuration in terms of the metric for the $k$ black hole configuration.
Notice that this relation precisely corresponds to the one found in \cite{Harmark:2003eg} from a more general point of view.

%%%%%%%%%%%%%%%%%%%%%%%%%%%%%%%%%%%%%%%%%%%%%%%%%%%%%%%%%%%%%%
\section{\label{sec:thermodynamics}Thermodynamics of the multi-black hole configuration}
%%%%%%%%%%%%%%%%%%%%%%%%%%%%%%%%%%%%%%%%%%

In this section we begin by determining the thermodynamic properties
of the multi-black hole configurations. This is accomplished in
Section \ref{cortherm1}. We subsequently find in Section
\ref{newtthermo} that the obtained thermodynamics is consistent with
a simple Newtonian interpretation.

%%%%%%%%%%%%%%%%%%%%%%%%%%%%%%%%%%%%%%%%%%%%%%%%%%%%%%%%%%%%%%
\subsection{Thermodynamic properties}
\label{cortherm1}
%%%%%%%%%%%%%%%%%%%%%%%%%%%%%%%%%%%%%%%%%%%%%%%%%%%%%%%%%%%%%%

In this section we find the thermodynamic quantities for multi-black
hole configurations on the cylinder to first order in the mass in
the limit of small total mass.

We begin by considering the quantities that one can read off from
the event horizons. For the $i^{\rm th}$ black hole the metric near
the horizon is given by \eqref{met1}-\eqref{met2}. The temperature
is now found in the standard way by computing the surface gravity
while the entropy is found from computing the area of the event
horizon divided with $4 G_{\rm N}$. This yields the following
entropy $S_i$ and temperature $T_i$ for the $i^{\rm th}$ black hole
\begin{eqnarray}
\label{corS} S_i = \nu_i^{\frac{d-1}{d-2}} \frac{\Omega_{d-1}}{4
G_{\rm N}} \rho_0^{d-1} \left( 1 + \frac{d-1}{d-2}\frac{
\Lambda^{(i)}}{2} \rho_0^{d-2} + \CO ( \rho_0^{2(d-2)} ) \right) \,,
\end{eqnarray}
\begin{eqnarray}
 \label{corT}
T_i = \nu_i^{-\frac{1}{d-2}} \frac{d-2}{4\pi \rho_0} \left( 1 -
\frac{d-1}{d-2}\frac{ \Lambda^{(i)}}{2} \rho_0^{d-2} + \CO (
\rho_0^{2(d-2)} ) \right) \,,
\end{eqnarray}
with $\Lambda^{(i)}$ as defined in (\ref{lambda}).

Turning to the asymptotic quantities, we need to determine the total
mass $M$ and the relative tension (binding energy) $n$. To determine
$M$ and $n$, we first notice the fact that the multi-black hole
solution obeys the first law of thermodynamics \cite{Harmark:2003eg}
\begin{equation}
\label{firstlaw} \delta M = \sum_{i=1}^k T_i \delta S_i \ .
\end{equation}
This is derived in \cite{Harmark:2003eg} using the ansatz
\eqref{ansatz} for a single connected horizon, but the argument
there is easily generalized to $k$ connected horizons. Note that in
\eqref{firstlaw} we do not have the variation of the circumference
of the cylinder since we have fixed the circumference to be $2\pi$.
This term is however easily added (see Ref.~\cite{Harmark:2003eg} and below).

It is a general property of the ansatz \eqref{ansatz} that $R_0$,
$M$ and $n$ are related as \cite{Harmark:2003eg}
\begin{equation}
\label{totalM1} M = \frac{\Omega_{d-2}}{8 G_{\rm N}} R_0^{d-3}
\frac{(d-1)(d-3)}{d-2-n} \,.
\end{equation}
This is easily seen from considering the metric \eqref{ansatz} for
$R \rightarrow \infty$. Using the definition of $\rho_0$ in
\eqref{defrho0} we can write this as
\begin{equation}
\label{totalM2} M = \frac{\Omega_{d-1}}{16\pi G_{\rm N}}
\rho_0^{d-2} \frac{(d-1)(d-2)}{d-2-n}\,.
\end{equation}
We can now insert \eqref{corS}, \eqref{corT} and \eqref{totalM2}
into the first law \eqref{firstlaw} for a given variation of
$\rho_0$, which yields the following result
\begin{equation}
\label{nrel} n + \frac{\rho_0}{d-2} \frac{\delta n}{\delta \rho_0} =
\frac{d-2}{2}  \sum_{i=1}^k \nu_i \Lambda^{(i)} \rho_0^{d-2} \ ,
\end{equation}
to first order in $\rho_0^{d-2}$. We used here that $n \rightarrow
0$ for $\rho_0 \rightarrow 0$. From  \eqref{totalM2} and
\eqref{nrel} we then conclude that $M$ and $n$, to first order in
$\rho_0^{d-2}$, are
\begin{eqnarray}
\label{corM}  M =  \frac{(d-1)\Omega_{d-1}}{16\pi G_{\rm N}}
\rho_0^{d-2} \left[ 1 + \frac{1}{4} \sum_{i=1}^k \nu_i \Lambda^{(i)}
\rho_0^{d-2}  + \CO ( \rho_0^{2(d-2)} ) \right] \ ,
\end{eqnarray}
\begin{eqnarray}
\label{firstn} n =  \frac{d-2}{4} \sum_{i=1}^k \nu_i \Lambda^{(i)}
\rho_0^{d-2} + \CO ( \rho_0^{2(d-2)} ) \ .
\end{eqnarray}
Thus, the physical quantities relevant for the thermodynamics of the
$k$ black hole configuration are given by \eqref{corS},
\eqref{corT}, \eqref{corM} and \eqref{firstn}.

We consider now how the relative tension $n$ and the entropies
$S_i$, as given above, behave as a function of the total mass $M$.
To this end, it is useful to define the rescaled mass $\mu$ as
\cite{Harmark:2003yz,Harmark:2005pp,Harmark:2007md}
\begin{equation}
\label{defmu} \mu \equiv \frac{16\pi G_{\rm N} M}{L^{d-2}} =
\frac{16\pi G_{\rm N} M}{(2\pi)^{d-2}}\\,
\end{equation}
where we used that the circumference $L=2\pi$. Using now
\eqref{corM} and \eqref{firstn}, we get that $n$ as function of
$\mu$ is given by
\begin{eqnarray}
\label{nofmu} n (\mu)  = \frac{(d-2)(2\pi)^{d-2} }{4(d-1) \Omega_{d-1}}
\sum_{i=1}^k \nu_i \Lambda^{(i)} \mu + \CO (\mu^2) \ .
\end{eqnarray}
We use this expression in Section \ref{sec:phasediag} since it gives
the linear slope of the multi-black hole configuration in the
$(\mu,n)$ phase diagram. There we also provide a rough estimate of
the range of $\mu$ for which \eqref{nofmu} is a good approximation.

Turning to the entropies, we have that the entropy of the $i^{\rm
th}$ black hole, in terms of the rescaled total mass $\mu$, is
\begin{eqnarray}
S_i (\mu) = \frac{ (2\pi)^{d-1} (\nu_i \,\mu)^{\frac{d-1}{d-2}} } { 4
 \Omega_{d-1}^{\frac{1}{d-2}} (d-1)^{\frac{d-1}{d-2}} G_{\rm N} }
 \left[ 1 + \frac{(2\pi)^{d-2}}{2(d-2)\Omega_{d-1}}\left( \Lambda^{(i)}
 -\frac{1}{2} \sum_{i=1}^k \nu_i \Lambda^{(i)} \right)
\mu + \CO ( \mu^2 ) \right] \,.
  \label{EntropyU}
\end{eqnarray}
One can now compute the total entropy $S_{\rm total} (\mu)$ as the
sum of the entropies \eqref{EntropyU} for each of the $k$ black
holes.

As already mentioned, the $k$ black hole configurations are unstable
with respect to small changes in the positions of the black holes.
Generic disturbances will destabilize the configuration and
presumably the $k$ black holes will merge into a single black hole.
Therefore, we expect in general that the entropy for a single black
hole is always greater than the total entropy of the $k$ black
holes, for same total mass $\mu$, \ie $S_{\rm total} (\mu) < S_{\rm
1BH}(\mu)$. This can indeed be verified from Eq.~\eqref{EntropyU},
for sufficiently small $\mu$. We examine these questions in detail
in Section \ref{sec:details2BH} for the two black hole case.

It is important to note that from the temperatures \eqref{corT} one
can see that they in general are not equal for the black holes in
the configuration. This means that generically the multi-black hole
configurations are not in thermal equilibrium. In fact, it is easy
to see from \eqref{corT} that the only configurations at this order that are in
thermal equilibrium are the copies of the single-black hole solution
studied previously in
\cite{Horowitz:2002dc,Harmark:2003eg,Harmark:2003yz}.

%%%%%%%%%%%%%%%%%%%%%%%%%%%%%%%%%%%%%%%%%%%%%%%%%%%%%%%%%%%%%%
\subsection{Newtonian interpretation of the thermodynamics} \label{newtthermo}
%%%%%%%%%%%%%%%%%%%%%%%%%%%%%%%%%%%%%%%%%%

The variable $\rho_0$ was useful to construct the multi-black hole
solution but is not the most appropriate one for the physical
interpretation of the solution and its thermodynamic quantities
(\ref{corS})-(\ref{firstn}), since it does not have an invariant
meaning.
 A more natural variable for the physical interpretation, as
will be confirmed below, is the ``areal'' radius. We define a set of
$k$ ``areal'' radii $\hat{\rho}_{0(i)}$, $i=1, \ldots, k$, by
\begin{equation}
\hat{\rho}_{0(i)} \equiv \nu_i^{\frac{1}{d-2}} \rho_0 \left( 1 +
\frac{\Lambda^{(i)}}{2(d-2)} \rho_0^{d-2} \right) \ .
\end{equation}
Using this definition the first-order corrected horizon area
of the $i^{\rm th}$ black hole takes the appropriate form
\begin{eqnarray}
{\cal A}_{\rm h}^{(i)}=
\Omega_{d-1} \hat{\rho}_{0(i)}^{d-1} \ ,
  \label{def:areal}
\end{eqnarray}
for a $(d-1)$-sphere of radius $\hat{\rho}_{0(i)}$.  We can now
rewrite, to leading order, the corrected thermodynamic quantities
(\ref{corS})-(\ref{firstn}) in terms of these ``areal'' radii.

The corrected entropy (\ref{corS}) and temperature \eqref{corT} of
the $i^{\rm th}$ black hole takes the form
\begin{equation}
\label{Si} S_i =  \frac{\Omega_{d-1}  \hat{\rho}_{0(i)}^{d-1}}{4
G_{\rm N}} \,, \qquad  T_i = T_{0(i)} ( 1 + \Phi_i )   \,, \qquad T
_{0(i)} \equiv \frac{d-2}{4\pi\hat{\rho}_{0(i)}} \ ,
\end{equation}
where we have defined the potential
\begin{equation}
\label{Phii}
\Phi_i = - \frac{\Lambda^{(i)}}{2} \rho_0^{d-2} \ .
\end{equation}
From the form of $\Lambda^{(i)}$ in Eq.~\eqref{lambda} we see that $\Phi_i$ is
precisely the Newtonian potential created by all images of
the  $i^{\rm th}$ black hole as well as all other $k-1$ masses (and
their images)  as seen from the location of the $i^{\rm th}$ black
hole. The interpretation of the form for the temperature in \eqref{Si} is that
$T_{0(i)}$ is the intrinsic temperature of the $i^{\rm th}$ black
hole, \ie when it would be isolated in flat empty $(d+1)$-dimensional space.
The second term  is the  redshift contribution coming from the gravitational potential
$\Phi_i$.

Similarly, the total mass \eqref{corM} of the configuration can be
written to leading order as
\begin{eqnarray}
\label{Mareal} M = \sum_{i=1}^k \left[ M_{0(i)} + \frac{1}{2}
M_{0(i)} \Phi_i\right]  \lspa\quad M_{0(i)} \equiv
\frac{(d-1)\Omega_{d-1}}{16\pi G_{\rm N}}\hat{\rho}_{0(i)}^{d-2} \ ,
\end{eqnarray}
where $\Phi_i$ is defined in \eqref{Phii}.
Again, the physical interpretation can be clarified as follows: The first term
\begin{equation}
M_{0} \equiv \sum_{i=1}^k M_{0(i)} \ ,
\end{equation}
is the some of the individual masses $M_{0(i)}$ when they would be isolated,
while the second term
\begin{equation}
\label{UNewton}
U_{\rm Newton}= \frac{1}{2}\sum_{i=1}^k  M_{0(i)} \Phi_i \ ,
\end{equation}
is precisely the negative gravitational (Newtonian) potential energy
that appears as a consequence of the black holes and their images.

From the above results it follows that one can derive the formula
for the relative tension in \eqref{firstn} by a purely Newtonian
argument, as was first done for the single black hole case in
Ref.~\cite{Gorbonos:2005px}. To see this, note that when we also
allow for the length $L$ of the circle to vary, the generalized
first law of thermodynamics \eqref{firstlaw} reads
\begin{equation}
\delta M = \sum_{i=1}^k T_i \delta S_i + \frac{n M }{L} \delta L \ ,
\end{equation}
since ${\cal{T}} = nM/L$ is the tension in the circle direction.
The relative tension can thus be computed from
\begin{equation}
\label{nder}
n = \frac{L}{M} \left( \frac{ \partial M}{\partial L} \right)_{S_i} \ .
\end{equation}
As described above, from a purely Newtonian analysis one knows that the total
mass $M= M_0 + U_{\rm Newton}$ is the sum of the intrinsic mass plus the
gravitational potential energy given in \eqref{UNewton}. Furthermore, the condition of keeping $S_i$ fixed
means that we should keep fixed the mass $M_{0(i)}$ of each black hole, and hence
also $M_0$. It thus follows from \eqref{nder} that to leading order
\begin{equation}
\label{nder2}
n = \frac{L}{M_0} \left( \frac{ \partial U_{\rm Newton}}{\partial L} \right)_{M_{0(i)}}
= -\frac{\rho_0^{d-2}}{4M_0} \sum_{i=1}^k M_{0(i)} L
\frac{ \partial \Lambda^{(i)}}{\partial L} \ ,
\end{equation}
where we used \eqref{UNewton}, \eqref{Phii} in the second step. To
compute the derivative we need to know how $\Lambda^{(i)}$ scales
with $L$. While the expression for $\Lambda^{(i)}$ in \eqref{lambda}
is for our choice $L=2\pi$, it is not difficult to see that keeping
$L$ arbitrary amounts to the rescaling $\Lambda^{(i)} \rightarrow
(2\pi/L)^{d-2} \Lambda^{(i)}$. Using this in \eqref{nder2} along
with $M_{0(i)}/M_0 = \nu_i$ immediately shows that we recover our
result \eqref{firstn} for the relative tension.

As a consequence, we conclude that the entire thermodynamics of the
first-order corrected multi black-hole solutions can be appropriately
interpreted from a Newtonian point of view.

%%%%%%%%%%%%%%%%%%%%%%%%%%%%%%%%%%%
\section{\label{sec:phasediag}Phase diagram for the multi-black hole configurations}
%%%%%%%%%%%%%%%%%%%%%%%%%%%%%%%%%%%

As mentioned in the Introduction, the whole set of different
multi-black hole configurations are part of a larger set of black
holes, black strings and other black objects which are
asymptotically $\CM^d \times S^1$
\cite{Harmark:2005pp,Harmark:2007md}. For this reason, it is very
useful to depict the multi-black hole configurations in a $(\mu,n)$
phase diagram \cite{Harmark:2003dg,Harmark:2003eg} in order to
understand the phase structure of all the solutions asymptoting to
$\CM^d \times S^1$.

\begin{figure}[h]
\begin{center}
\resizebox{10cm}{8cm}
 {\includegraphics{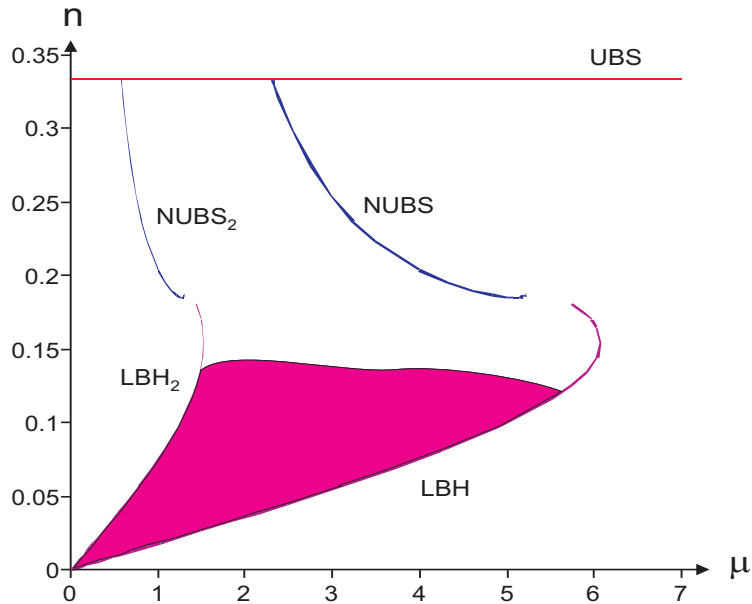}}
\end{center}
\caption{Phase diagram for $d=5$ with $n$ versus $\mu$ for the
two-black hole configurations spanning the area in between the
single black hole (LBH) and two equal size black holes (LBH$_2$).
Moreover, we have drawn the uniform black string phase (UBS), the
non-uniform black string phase (NUBS) and its two-copied phase
(NUBS$_2$).}
 \label{fig:phasediag}
\end{figure}

A multi-black hole configuration corresponds to a point in the
$(\mu,n)$ phase diagram. The coordinates of this point are given by
\eqref{corM}-\eqref{defmu}. However, since we look at the limit of
small gravitational interactions, it is useful to have $n$ as function of $\mu$.
This is given by \eqref{nofmu}. Therefore, $n(\mu)$ as given in
\eqref{nofmu} is valid for small $\mu$. For a fixed $\mu$, one can
then consider the range of $n$ for a configuration with $k$ black
holes. This can be seen using the following inequality for a $k$
black hole configuration
\begin{eqnarray}
\label{range n} \frac{2\zeta (d-2)}{(2 \pi)^{d-2}}< \sum_{i=1}^k
\nu_i \Lambda^{(i)} \leq k^{d-3} \frac{2\zeta (d-2)}{(2 \pi)^{d-2}}
\ ,
\end{eqnarray}
where $\Lambda^{(i)}$ is defined in \eqref{lambda}. The lower bound
corresponds here to the single-black hole case ($k=1$) while the
upper bound corresponds to the case of $k$ black holes of equal
mass, distributed equidistantly around the cylinder. These are the
so-called copies of the single black hole on the cylinder considered
in \cite{Horowitz:2002dc,Harmark:2003eg,Harmark:2003yz}. Now, using
the inequality \eqref{range n} with \eqref{nofmu} we see that in the
$(\mu,n)$ phase diagram the $k$ black hole configurations
corresponds to the points lying above the single-black hole phase
and below the $k$ copied phase. We have depicted this for $d=5$ in
Fig.~\ref{fig:phasediag} in the case of two black holes on the
cylinder ($k=2$). We see that the phases with two unequal black
holes lie in between the single localized black hole phase (LBH)
and the phase with two equal size black holes (LBH$_2$). We have
depicted here the phases using the numerical data found in
\cite{Kudoh:2004hs} for the single black hole phase (LBH). Note that
it is not clear that the phases of the two black hole configurations
will stay in between the LBH and LBH$_2$ phases when we go beyond
our perturbative solution (see discussion in Section \ref{sec:phasestruc}).
In Fig.~\ref{fig:phasediag} we have
furthermore depicted the uniform black string phase (UBS), which has
$n=1/(d-2)$, and the non-uniform black string phase (NUBS), along
with the two-copied non-uniform black string phase (NUBS$_2$).%
\footnote{For the $d=5$ non-uniform black string we have used the
data given in \cite{Wiseman:2002zc,Harmark:2003dg}. The map to the
two-copied solution is given in \cite{Harmark:2003eg}.}

From \eqref{nofmu} and the inequality \eqref{range n} we see that
for a given mass $\mu$ we have a continuously infinite
non-uniqueness of solutions with $k$ black holes. However, the
non-uniqueness of solutions is even worse than this. If we consider
a $k$ black hole solution it is described by $k$ parameters, as
explained in Section \ref{sec:EquilSolutions}. Thus, since the
solutions with $k$ black holes span a two-dimensional area in the
$(\mu,n)$ diagram we need $k-2$ extra parameters, beyond $\mu$ and
$n$, to point to a specific solution with $k$ black holes.
Therefore, there is a continuous infinite non-uniqueness of
solutions for certain points in the phase diagram, when $k \geq 3$.
Moreover, if we do not specify $k$ but instead consider all possible
multi-black hole configurations, we have an infinite layer of
solutions in the phase diagram, since one can always consider adding
a small black hole to a given multi-black hole configuration.

Hence our results show a continuous non-uniqueness for solutions with fixed $M$.
Such non-uniqueness was also observed in Ref.~\cite{Elvang:2004iz} for
bubble-black hole sequences, which are not spherically symmetric on $ \R^{d-1}$
and lie in the region $ \frac{1}{d-2} \leq n \leq d-2$ of the $(\mu,n)$ phase diagram.
The multi-black hole configurations of this paper are therefore the first example
of continuous non-uniqueness for solutions that are spherically symmetric on $ \R^{d-1}$.

Considering the phase diagram for the two black hole configurations
depicted in Fig.~\ref{fig:phasediag}, it is interesting to
consider what happens when moving up in $n$. One way to do this is
to increase $\mu$ such that the ratios $\nu_i$ are fixed. In this
case the two black holes are growing and eventually their horizons
will meet. Thus, the natural question is then what happens when
approaching this point. There seems to be two possibilities:
\begin{itemize}
\item[1)] When the horizons of the two black holes meet, their
temperatures are not equal, and the solution will be singular in the
meeting point.
\item[2)] The temperatures of the two black holes will approach each
other and when the two black holes meet they will merge into a new
non-uniform black string phase different from both the original
non-uniform black string phase emanating from the Gregory-Laflamme
point, and the two-copied non-uniform black string phase.
\end{itemize}
We explore these scenarios further in Section \ref{sec:details2BH}.
In Section \ref{sec:conclusions} we discuss the possible
implications for the Gregory-Laflamme instability if there should
exist new non-uniform string phases.

Finally, we note that it is useful to give a rough estimate of the
validity of the perturbative $k$ black hole solutions found in
Section \ref{sec:construction}. For this purpose we can employ the
estimate made for the single black hole solution in
\cite{Harmark:2003yz}. A lower estimate can be found by considering
the $k$ copied phase, since we expect this to be the first solution
for which the first order correction becomes invalid, as one
increases $\mu$.  We therefore take the function $F(\rho,\theta)$ in
\eqref{lim F} and consider when the contribution from the term with
$\Lambda^{(i)}_2$ is equal to the one with $\Lambda^{(i)}$. This
happpens for $\rho^2 \simeq 8\pi^2 \zeta (d-2)/(k^2
(d-1)(d-2)\zeta(d))$. This can be used to get an upper bound for the
Schwarzschild radius $k^{-1/(d-2)} \rho_0$. Plugging that into $\mu$
in terms of $\rho_0$, one obtains a rough upper bound on $\mu$. For
$k$ black holes, this means that the method is valid in the regime
$\mu \ll \mu_*$, with $\mu_* = 30/k, 9/k^2, 1.8/k^3, 0.2/k^4,
0.02/k^5, 0.002/k^6$ for $d=4,5,6,7,8,9$. Therefore, for $k=2$ and
$d=5$ we get that our perturbative solutions describing two black
holes on the cylinder are valid for $\mu \ll 2.2$, in accordance
with Fig. \ref{fig:phasediag}. The values $\mu_*$ for $k=2$ black
hole copies in $4\leq d \leq 9$ will be given in Table \ref{tabmuc}
in Section \ref{sec:details2BH}.

%%%%%%%%%%%%%%%%%%%%%%%%%%%%%%%%%%%
\section{\label{sec:details23bh}Further analysis of specific solutions}
%%%%%%%%%%%%%%%%%%%%%%%%%%%%%%%%%%%

In this section we analyze in more detail the two simplest
multi-black hole configurations, namely two- and three-black hole
solutions. This serves as an illustration of the general solution
and its physical properties, but will also provide us with further
insights into the structure of the phase diagram discussed in the
previous section, including the possibility of existence of new
lumpy black holes in Kaluza-Klein spaces.

%%%%%%%%%%%%%%%%%%%%%%%%%%%%%%%%%%%
\subsection{\label{sec:details2BH}Two-black holes on the cylinder}
%%%%%%%%%%%%%%%%%%%%%%%%%%%%%%%%%%%

We start by examining the case of the two-black hole solution, \ie
we take a configuration of two black holes with mass fractions
$\nu_1 = \frac{1}{2} + \kappa$ and $\nu_2 = \frac{1}{2} - \kappa$, where $ 0 \leq \kappa
\leq 1/2$ so that by convention $M_1 \geq M_2$. Hence, $\kappa =0$
corresponds to a configuration with two black holes of equal mass,
while the limiting case $\kappa = 1/2$ is the single black hole
solution. The locations of the black holes are chosen as $z_1^* = 0$
and the location of the second black hole is denoted as $z_2^*$. For
the equilibrium configuration we clearly have $z_2^* = \pi$ so that
the two black holes are on opposite points on the circle.

We first focus on the equilibrium configuration. To compute the
various thermodynamic quantities we need $\Lambda^{(1,2)}$ defined
in \eqref{lambda}, which are given explicitly for the two-black hole
case in Eq.~\eqref{lambda2BHs0}. Furthermore, the expression for the
sum $\sum_{i=1}^2 \nu_i \Lambda^{(i)}$ is given in \eqref{sum2bh}.
 The curve \eqref{defmu} in the phase diagram is thus given by
\begin{equation}
 n  (\mu;\kappa) = \frac{ (d-2) \zeta (d-2)}{ (d-1) \Omega_{d-1}} 2^{d-4}
 \left[ 1 - 4 \kappa^2 \Big( 1 - 2^{3-d} \Big)  \right] \mu + {\cal{O}} (\mu^2) \ .
\end{equation}
Since the constant of proportionality is a monotonically increasing
function of $\kappa$ one sees here explicitly that the inequality
\eqref{range n} at $k=2$ is obeyed, so that the slope in the
$(\mu,n)$ phase diagram is bounded by that of a single black hole
and two equal mass black holes.

\begin{figure}[h]
\begin{center}
\resizebox{6cm}{4cm}
 {\includegraphics{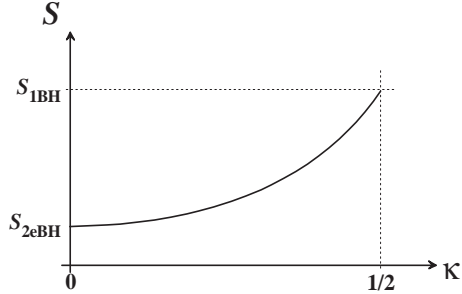}}
\end{center}
\caption{Plot of the total entropy $S$ of an equilibrium two-black
hole configuration as a function of its mass distribution $\kappa$,
for a fixed total mass $\mu$. This is a schematic plot for $\mu
<\mu_{\rm c}$.} \label{fig:Splot1}
\end{figure}

From \eqref{EntropyU} and \eqref{lambda2BHs0} we find the total
entropy is given by
\begin{eqnarray}
\label{2bhentro}
S (\mu;\kappa) & =  & S_1( \mu ; \kappa) + S_1 (\mu; -\kappa) \ , \\
S_1 (\mu;\kappa)  & = & \frac{ (2\pi)^{d-1}  } { 4
 \Omega_{d-1}^{\frac{1}{d-2}} (d-1)^{\frac{d-1}{d-2}} G_{\rm N} }
\left[ \Big( \frac{1}{2} + \kappa \Big) \,\mu \right]^{\frac{d-1}{d-2}} \nonumber \\
 && \times
\left[ 1 + \frac{\zeta (d-2)}{(d-2) \Omega_{d-1} } \left( 2^{d-4} +
2 \kappa (1 - \kappa) (1 - 2^{d-3})
 \right) \mu + {\cal{O}} (\mu^2)  \right] \ , \label{S1cor}
\end{eqnarray}
where we used that $S_2 (\mu;\kappa) = S_1 (\mu;-\kappa)$. In
particular, we find from this the corrected entropy of one black
hole on a circle $S_{\rm 1BH} (\mu)  \equiv S (\mu;1/2)$ and that of
two equal mass black holes $S_{\rm 2eBH} (\mu) \equiv S (\mu;0)$. We
can now consider $S (\mu;\kappa)$ for fixed total (rescaled) mass
$\mu$ as
 $\kappa$ ranges from 0 to $1/2$. Physically, we expect that this
is a monotonically increasing function of $\kappa$ since it should be
entropically favored to have all the mass concentrated in one black
hole, and the solution with two black holes is in an unstable
equilibrium. As shown in Fig.~\ref{fig:Splot1}, this is indeed the
behavior we find when the mass of the system is not too large.

We can in fact use the physical criterion that $S(\mu;\kappa)$ be a
monotonically increasing function of $\kappa$ to get an upper bound
$\mu_{\rm c}$ on the mass, in order for our perturbative approach to
be valid. By examining the function \eqref{2bhentro} in detail, we find that
a condition that can be used to determine the critical mass is
\begin{eqnarray}
\frac{\partial^2 S (\mu=\mu_{\rm c};\kappa)}{\partial \kappa^2}
{\biggr |}_{\kappa=0}=0 \, , \label{condMcrit}
\end{eqnarray}
where above we also used that $ (\partial S(\mu,\kappa)/\partial \kappa ) \vert_{\kappa =0}
=0 $ for all $\mu$. Using the explicit expression \eqref{2bhentro}
we have analyzed this equation for $ 4 \leq d \leq 9$ and the
results for $\mu_{\rm c}$ are listed in Table \ref{tabmuc}. A
necessary condition for our method to be valid is thus $\mu \lesssim
\mu_{\rm c}$. We expect that for $\mu \ll \mu_{\rm c}$ our
perturbative solution for the two-black hole configuration is valid.
As illustrated in Table \ref{tabmuc}, this is a less restrictive
bound than the one found in the end of Section \ref{sec:phasediag}
based on a less precise consideration.

\begin{table}[ht]
\begin{center}
\begin{tabular}{|c||c|c|c|c|c|c|}
\hline $d$ & $4$ & $5$ & $6$ & $7$ & $8$ & $9$ \\ \hline \hline
$\mu_{\rm c}$ & 14.4 & 7.1 & 3.9 &2.0 & 0.97 & 0.44 \\ \hline
$\mu_{\rm *}$ & 15 & 2.2 & 0.2 & $1\times 10^{-2}$ & $8\times 10^{-4}$ & $3\times 10^{-5}$ \\
\hline $\mu_{\rm GL}$ & $3.52$ & $2.31$ & $1.74$ & $1.19$ & $0.79$ &
$0.55$
 \\
\hline
\end{tabular}
\caption{The upper bound $\mu_{\rm c}$, imposed by entropy
arguments, on the mass for the validity of the perturbative
two-black hole results. For comparison the bound $\mu_*$ (see end of
Section \ref{sec:phasediag}) is shown along with the
Gregory-Laflamme masses $\mu_{\rm GL}$ (see e.g.
\cite{Harmark:2003dg}). \label{tabmuc}}
\end{center}
\end{table}

It is also useful to examine the temperatures of each of the black
holes as we increase the mass. Clearly, for two black holes of
unequal mass the zeroth order temperatures are different, and the
system is not in thermal equilibrium. However, we can calculate the
effect of the redshift on the ratio of temperatures, and examine
whether this effect tends to equilibrate the black holes as we increase the
total mass of the system. Using the first-order corrected
temperatures \eqref{corT} and the expressions \eqref{lambda2BHs0}
for $\Lambda^{(1,2)}$ one finds
\begin{eqnarray}
\label{ratioTemp} \frac{T_2}{T_1} = \left( \frac{1 +  2 \kappa}{1 -
2 \kappa} \right)^{\frac{1}{d-2}} \left[  1 -   \frac{4 \, \kappa \,
\zeta (d-2) }{ (d-2) \Omega_{d-1} } (2^{d-3} -1)  \mu
 + {\cal{O}} (\mu^2) \right] \ ,
\end{eqnarray}
where we eliminated $\rho_0$ in favor of $\mu$ using \eqref{corM},
\eqref{defmu}.

For two unequal mass black holes (with $M_1 > M_2$) we have $0 <
\kappa < 1/2$ so the pre-factor in \eqref{ratioTemp} is greater than
one. We now observe that as one increases the total mass $\mu$
the linear factor in $\mu$ will be smaller than one, causing the
ratio $T_2/T_1$ to decrease towards one. We thus conclude that the
first order redshifts combine in such a way that increasing the
total mass of two unequal mass black holes causes the temperatures
of the two black holes to approach each other. This indicates that
it may be possible that in the full non-perturbative regime the
temperatures converge to a common value at the merger.

\begin{figure}[h]
\begin{center}
\resizebox{6cm}{4cm}
 {\includegraphics{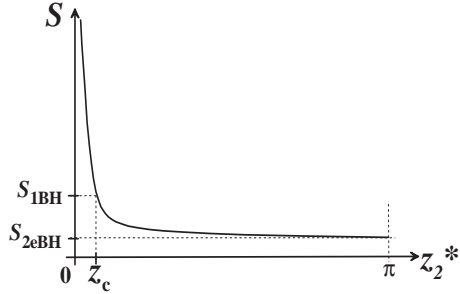}}
\end{center}
\caption{Plot of the total entropy $S$ of a two black hole
configuration with fixed total mass $\mu$ and fixed mass
distribution (here $\kappa =0$) as a function of the relative
distance $z_2^*$ between the two black holes. We use a values of
$\mu$ that lies below the critical mass $\mu_{\rm c}$ listed in
Table \ref{tabmuc}.}
 \label{fig:SplotZ}
\end{figure}

Finally, we study the entropy of the more general configuration of
two black holes without requiring the system to be in mechanical equilibrium.
The total entropy $S(\mu;\kappa,z_2^*)$ is obtained by using again
\eqref{EntropyU} to compute the individual entropies $S_{1,2}$,
 but now substituting the $z_2^*$-dependent functions $\Lambda^{(1,2)} (z_2^*)$ given in
 Eq.~\eqref{lambda2BHs}. We consider then  a fixed total mass $\mu$ and
 mass distribution $\kappa$, and vary the location $z_2^*$ of the second black hole
where $0 < z_2^*\leq \pi$. Physically we expect that
$S(\mu;\kappa,z_2^*)$ is a monotonically decreasing function of
$z_2^*$, with minimal entropy when the black holes are farthest
apart and maximal entropy when they have merged into a single black
hole. This is indeed the case, as shown in Fig.~\ref{fig:SplotZ}.

We can view the decrease of  $z_2^*$ as a time evolution process in
which two black holes initially separated by a distance $\pi$ on the
circle are perturbed and then collapse into a single black hole. As
seen in Fig.~\ref{fig:SplotZ} the total entropy increases during
this process, but the entropy diverges as the distance between the
black holes goes to zero. This is expected since fields diverge when
we let the distance between sources go to zero in the point-particle
limit, and indeed $\Lambda^{(1,2)} (z_2^*)$ in \eqref{lambda2BHs}
diverge as $z_2^*\rightarrow 0$. However, for physical sources, the
minimum distance of approach between the sources is given by their
size. In our case, a good estimate for this critical distance is
given by the horizon size of a $(d+1)$-dimensional Schwarzschild
(spherical) black hole with total mass $\mu$, given by
\begin{equation}
\label{rhos} \rho_{\rm s} \equiv 2\pi  \left(
\frac{\mu}{(d-1)\Omega_{d-1} } \right)^{\frac{1}{d-2}}\,.
\end{equation}
On the other hand, we can compute
the distance $z_{\rm c}$ at which the entropy curve
$S(\mu;\kappa,z_2^*)$ crosses the entropy $S_{\rm 1BH} (\mu) $ of a
single black hole configuration (see Fig.~\ref{fig:SplotZ}), \ie
\begin{equation}
S(\mu;\kappa,z_{\rm c}) = S_{\rm 1BH} (\mu) \ .
\end{equation}
Comparison of the two critical distances $\rho_{\rm s}$ and $z_{\rm c}$
now provides an important check on the validity of our perturbative
method, since we expect these two numbers to be of the same order.
As illustrated in Table \ref{tab} this match indeed occurs with
$\rho_{\rm s} > z_{\rm c}$, where for definiteness we have chosen
$\kappa =0$.

We thus conclude that also for non-equilibrium configurations the
corrected thermodynamics leads to physically sensible results.

\begin{table}
\begin{center}
\begin{tabular}{|c||c|c|c|c|c|c|c|c|c|c|c|c|}
\hline
$d$ & \multicolumn{4}{c|}{4} & \multicolumn{4}{c|}{5} & \multicolumn{4}{c|}{6} \\
\hline \hline $\mu $  & 0.01 & 0.1 & 1 & 10 & 0.01 & 0.1 & 1 & 7 &
0.01 & 0.1 & 1 & 3   \\ \hline $z_{\rm c}$ & $~0.055$ & $~0.17$&
$~0.52$& $~1.26$& $~0.25$& $~0.53$& $~1.13$ & $~1.97$& $~0.54$&
$~0.95$& $~1.68$& $~2.25$ \\ \hline $ \rho_{\rm s}$ & $~0.081$& $~0.26$&
$~0.82$ & $~2.58$ & $~0.29$& $~0.62$& $~1.33$ & $~2.54$& $~0.56$&
$~1.00$& $~1.78$& $~2.34$ \\ \hline
\end{tabular}
\end{center}
\caption{\label{tab} Comparison of the two critical distances
$z_{\rm c}$ and $\rho_{\rm s}$ in the case $\kappa =0$ for some
representative values of $d$ and $\mu$ (taken below the mass
$\mu_{\rm c}$  for which we can trust the perturbative results, see
Table \ref{tabmuc}). $z_{\rm c}$ is the minimum distance imposed by
entropic considerations, as illustrated in Fig.~\ref{fig:SplotZ},
and $\rho_{\rm s}$ is the size of a $(d+1)$-dimensional Schwarzschild
black hole with mass $\mu$. }
\end{table}

%%%%%%%%%%%%%%%%%%%%%%%%%%%%%%%%%%%
\subsection{\label{sec:details3BH}Three-black holes on the cylinder}
%%%%%%%%%%%%%%%%%%%%%%%%%%%%%%%%%%%

In this subsection we discuss some features that can be addressed
when we have three (or more) black holes, and we skip properties
that are already present in the two-black hole configuration. In particular,
by studying merges of two black holes we find evidence for new ``lumpy'' black
hole configurations.

For definiteness, take a symmetric three-black hole configuration in
equilibrium, located at the points $z_1^*=0$, $z_2^*=\pi-y$, and
$z_3^*=\pi+y$. We also adjust the masses $M_i=\nu_i M$ such that
$\nu_2=\nu_3=\frac{1}{2}(1-\nu_1)$, \ie black hole 2 and 3 have
equal mass. We now want to increase the total mass of the system
while maintaining equilibrium. The black holes will thus increase in
size and fill more and more of the free space in between them. The
question we want to address is whether the two black holes 2 and 3
with the same mass will merge first, before merging with black hole
1, or whether black hole 1 will merge with the other two before 2
and 3 can merge.

As in the previous subsection, our answer to this question is
limited by the fact that our formulae are  strictly valid only for
small black holes interacting via Newtonian gravity, while the black
hole merging process we wish to consider is certainly one where the
full nonlinearities of Einstein's equations are important. However,
we expect that with the available construction we can gain useful
insights into the behavior of the system, so we proceed to examine
this situation keeping in mind potential caveats.

The question above can be addressed by analyzing the ratio
\begin{eqnarray}
\label{ratio} X = \frac{\rho_{{\rm s}(1)}  + \rho_{{\rm s}(2)}}{z_{12} }
\frac{z_{23} } {\rho_{{\rm s}(2)}  + \rho_{{\rm s}(3)} } \ ,
\end{eqnarray}
where $ \rho_{{\rm s}(i)}$ is the Schwarzschild radius of the $i^{\rm th}$
black hole (defined as in \eqref{rhos}) and $z_{ij}$ is the distance
between the $i^{\rm th}$ and $j^{\rm th}$ black holes. It is not
difficult to see that this ratio is appropriate. Indeed, if black
hole 1 joins 2 (and 3, by symmetry) first then at the point they
merge
 one has $ \frac{z_{12}}{\rho_{{\rm s}(1)}  + \rho_{{\rm s}(2)}} =1$ and
$\frac{z_{23}}{\rho_{{\rm s}(2)}  + \rho_{{\rm s}(2)} }>1$, so that  $X>1$. On
the other hand, if 2 and 3 merge first then at one has $X<1$ at the
merging point.

We can express the ratio $X$ defined in \eqref{ratio} as a function
of the distance $y$ between black hole 1 and 2 (and 3) as follows.
First one uses the relation $\rho_{{\rm s}(i)}\propto \left(\nu_i
M\right)^{\frac{1}{d-2}}$ between the Schwarzschild radius and the
black hole mass in $d+1$ dimensions along with the fact that $\nu_2
= \nu_3$, so that
\begin{equation}
\label{rhoration} \frac{\rho_{{\rm s}(1)}  + \rho_{{\rm s}(2)}}{\rho_{{\rm s}(2)}  +
\rho_{{\rm s}(3)} } =
\frac{1}{2}\left[1+\left(\frac{\nu_1}{\nu_2}\right)^{\frac{1}{d-2}}\right]
=
\frac{1}{2}\left[1+\left(\frac{V_{23}}{V_{12}}\right)^{\frac{1}{d-2}}\right]
\ ,
\end{equation}
where we used the equilibrium conditions \eqref{fromnoforce} in the
last step. Finally, we substitute the explicit expressions
(\ref{Vij}) for $V_{ij}$ where $z_{12}=z_2^*-z_1^* =\pi-y $ in
$V_{12}$ and $z_{23}=z_3^*-z_2^*=2y $ in $V_{23}$. Note that
equilibrium requires $z_{23}<\pi$, so we only consider $0 < y <
\pi/2$. Equilibrium also demands that $\nu_2=\nu_3<\nu_1/2$.

Collecting results, we use \eqref{rhoration} to write the ratio in \eqref{ratio}
as
\begin{eqnarray}
\label{ratio3} X (y)  &=&\frac{y}{\pi-y} {\biggl [} 1+
\left(\frac{\pi-y}{2y}\right)^{\frac{d-1}{d-2}}
  \\
&&\qquad \quad \times
 \left( \frac{ (2 \pi)^{d-1} -(2y)^{d-1} \left [ \zeta
\left(d-1,1-\frac{y}{\pi}\right) - \zeta
\left(d-1,1+\frac{y}{\pi}\right) \right] }
 {(2 \pi)^{d-1} -(\pi-y)^{d-1} \left [ \zeta
\left(d-1,1-\frac{\pi-y}{2\pi}\right) - \zeta
\left(d-1,1+\frac{\pi-y}{2\pi}\right) \right]}
\right)^{\frac{1}{d-2}}{\biggr ]}. \nonumber
\end{eqnarray}
We can understand (\ref{ratio3}) as follows. A given value of $y$ fixes
 the location and mass distribution of the system in
equilibrium. Now let the total mass of the system increase. There is
a critical value, call it $M_{23}$, above which 2  and 3  are
merged. Similarly above a critical value $M_{12}$, 1 is merged
with 2 (and 3). $X$ can then be expressed as the ratio
$\left(M_{23}/M_{12}\right)^{\frac{1}{d-2}}$. Thus, if $X<1$, as we
increase the total mass of the distribution, black hole 2 and 3 will
merge before 1 joins them, and vice-versa.

\begin{figure}[h]
\begin{center}
\resizebox{6cm}{4cm}
 {\includegraphics{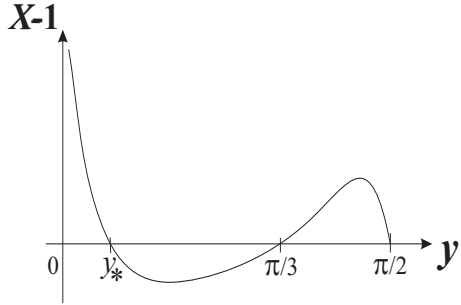}}
\end{center}
\caption{A typical plot of $X-1$ versus the distance $y$ ranging from 0 to $\pi/2$.}
\label{figX}
\end{figure}

A numerical analysis of (\ref{ratio3}) shows the following features,
see Fig.~\ref{figX}.
For $\frac{\pi}{3}<y<\frac{\pi}{2}$, one has $X>1$; for
$y_{*}<y<\frac{\pi}{3}$, $X<1$; and for $0<y<y_{*}$, $X>1$ again.
Here, $y_{*}$ is a critical value that depends on the dimension of
the spacetime. For example, for $d=4$ one has $y_{*}\sim
\frac{\pi}{3.76}$, while for $d=9$ one has $y_{*}\sim
\frac{\pi}{55.56}$. More generally, as $d$ grows $y_{*}$ decreases  and the interval
where $X <1$ grows.

To interpret these results first note that the case  $y=\pi/2$
describes a two-black hole configuration that is the limiting case of
the three-black hole configuration where $M_1\rightarrow 0$, and we
have two equal black holes each with mass $M/2$ located at
$z=\pi/2$ and $z=3\pi/2$. The case $y=\pi/3$ corresponds to a
symmetric configuration with three equal black holes equally spaced
along the circle. The case $y=0$ yields essentially the single-black hole
limit of the three-black hole configuration where $M_1=M$ is
centered at $z=0$ while $M_2 = M_3 \rightarrow 0$. Therefore, as $y$ goes from $0$ to
$\pi/2$, the masses $M_2=M_3$ increase from $0$ up to $M/2$, while
$M_1$ decreases from $M$ to $0$.

Keeping these features in mind, and that
$\frac{\pi}{3}<y<\frac{\pi}{2}$ implies $z_{23}>z_{12}$, it follows
that as the total mass increases black hole 2 and 3 will merge with
black hole 1 before they meet each other. We thus expect, as
observed above, that $X>1$ for these values of $y$ and $X=1$ at the
boundaries of the range. To understand the behavior of $X$ for
$0<y<\frac{\pi}{3}$ it does not suffice to use purely geometrical
arguments. Indeed, since $M_1>M_2$ in this branch, black hole 1
seems to approach 2 (and 3) faster than 2 and 3 approach each other,
but since at the same time $z_{23}<z_{12}$, we should use the
numerical analysis of $X$ described above to determine what happens.
This tells us that black hole 2 and 3 merge first, at least for
$y_{*} <y<\frac{\pi}{3}$ where we found $X<1$. However, for $y$
values smaller than  $y_{*}$ the numerical results for $X$ are not
reliable anymore, since in particular as $y \rightarrow 0$ we see
that $X\rightarrow \infty$ which is due to the fact that our
formulae are strictly valid in the point-particle limit where fields
diverge when the distance between sources vanishes.

The results above suggest that it could be possible that after the
merging of the two black holes (2 and 3) we end up with a``lumpy"
black hole (\ie a `peanut-like' shaped black object) together with
an ellipsoidal  black hole (1).
It is conceivable that such a configuration would be a new static
black hole solution in asymptotically $\CM^d \times S^1$ spacetimes.
Generally if two black holes were to merge in this way,
we expect that the resulting configuration would be singular. The singularity
would arise if the surface gravities or temperatures of the two black holes differed,
following standard results of \cite{Bardeen,Racz}. In the above construction, however, we chose
$M_2 = M_3$ to make the surface gravities identical in the merger.

To discuss this further, note first of all that it is still true
that the area of one spherical black hole of given mass is bigger
than the sum of the areas of two isolated black holes with the same
total mass. Nevertheless, the following argument suggests the
possibility of lumpy objects for $d\geq 4$. In general dimension
$d+1$, the horizon radius of a Schwarzschild black hole scales as
$\rho_{{\rm s}}\sim M^{\frac{1}{d-2}}$, so starting from two black
holes with $\rho_{{\rm s}(1,2)}\propto M_{1,2}^{\frac{1}{d-2}}$ we
have at the merging point a total radius $\rho_{{\rm
s}(1)}+\rho_{{\rm s}(2)}\propto
M_{1}^{\frac{1}{d-2}}+M_{2}^{\frac{1}{d-2}}$. On the other hand a
single black hole with mass $M_1+M_2$ has a radius $\rho_{{\rm
s}(12)}\propto (M_{1}+M_2)^{\frac{1}{d-2}}$. In four dimensions
($d=3$) this scales the same way as the total radius of the merged
object so we expect the formation of a spherical black hole
\cite{Thorne}. However, for $d \geq 4$ the power in the exponent is
less than one so that $\rho_{{\rm s}(1)}+\rho_{{\rm s}(2)} >
\rho_{{\rm s}(12)}$. Hence we should expect that the resulting
merger configuration will not be surrounded by a spherical horizon,
as would occur for $d=3$. % v2

As a consequence, it seems that for $d\geq 4$ this analysis does not
rule out the possibility of having a configuration of a lumpy black
object with ``centers" at $z=\pi\pm y$ kept in an unstable
equilibrium by a black hole at $z=0$ (and the respective copies).
Note also that the argument above suggests that the higher the
dimension, the more likely it is that lumpy black holes will exist.
Finally, we emphasize that the above analysis should be read within
the earlier-mentioned limitations of our construction.

Note that in asymptotically flat space new stationary black holes with similar
`rippled' horizons  of spherical topology
have been argued to exist in Ref.~\cite{Emparan:2003sy} by
considering ultraspinning Myers-Perry black holes in dimensions
greater than six. While in that case the ripples are supported by the angular momentum
$J$ in our case they are supported by the external stress of the other
(ellipsoidal) black hole. It would be interesting to generalize the analysis above
to configurations with more bumps, for example taking a symmetric four-black hole
configuration with $M_2 = M_3 = M_4$.

%%%%%%%%%%%%%%%%%%%%%%%%%%%%%%%%%%%
\section{\label{sec:conclusions}Conclusions and outlook}
%%%%%%%%%%%%%%%%%%%%%%%%%%%%%%%%%%%

%%%%%%%%%%%%%%%%%%%%%%%%%%%%%%%%%%%
\subsection{Summary}
\label{sec:sum}
%%%%%%%%%%%%%%%%%%%%%%%%%%%%%%%%%%%

In this paper we constructed solutions of the vacuum Einstein equations
describing multi-black hole configurations on the cylinder $\R^{d-1} \times S^1$ with $d\geq 4$, in the
limit of small total mass, or, equivalently, in the limit of a large cylinder. These solutions generalize the analytic
solutions found for the single black hole on the cylinder
\cite{Harmark:2003yz,Gorbonos:2004uc,Gorbonos:2005px,Karasik:2004ds,Chu:2006ce}.
Furthermore, they generalize the so-called copies of the single
black hole solutions corresponding to having equal mass black holes
distributed equidistantly around the cylinder
\cite{Horowitz:2002dc,Harmark:2003eg}. The new solutions are valid
to first order in the total mass, and are constructed using the technique of
\cite{Harmark:2003yz} based on an ansatz for the metric found in
\cite{Harmark:2002tr}.

Using the first-order corrected metrics for the multi-black hole
configurations we have studied their thermodynamics. Included in this
is one of the central results of this paper: The relative tension
(binding energy) $n$ as a function of the total (rescaled) mass
$\mu$, as given by Eq.~\eqref{defmu}. Using this, we have shown how
the solutions appear in the $(\mu ,n)$ phase diagram
\cite{Harmark:2003dg,Harmark:2003eg}, together with the other known
solutions that asymptote to $\CM^d \times S^1$. We observed that a
multi-black hole configuration with $k$ black holes has $k$
independent parameters. This implies a continuous non-uniqueness in
the $(\mu,n)$ phase diagram (or for a given mass), much like the one
observed for bubble-black hole sequences \cite{Elvang:2004iz}.

 The multi-black
hole configurations have to be in mechanical equilibrium in order to have a
static solution. We have identified where this requirement appears
in the construction of the solution, and we have furthermore
examined how to build such equilibrium configurations. Moreover, we have
described a general copying mechanism that enables us to build new
equilibrium configurations by copying any given equilibrium
configuration a number of times around the cylinder.

Finally, we examined in detail configurations with two and three
black  holes. For two black holes we verified the expectation that
one maximizes the entropy by transferring all the mass to one of the
black holes, and also that if the two black holes are not in mechanical
equilibrium then the entropy is increasing as the black holes become
closer to each other. These two facts are both in accordance with
the general argument that the multi-black hole configurations are in
an unstable equilibrium and generic perturbations of one of the positions
will result in that all the black holes merge together in a single
black hole on the cylinder. For the three black hole solution we
examined and found preliminary evidence for the hypothesis that for
certain three-black hole configurations two of the black holes can
merge into a lumpy black hole, where the non-uniformities are
supported by the gravitational stresses imposed by an external field.

From the first-order corrected temperatures one can show that the
multi-black hole configuration are in general not in thermal
equilibrium. The only configurations that are in thermal equilibrium
to this order are the copies of the single-black hole solution
studied previously \cite{Horowitz:2002dc,Harmark:2003eg,Harmark:2003yz}.
As a further comment we note that Hawking radiation will seed the mechanical
instabilities of the multi-black hole configurations.
The reason for this is that in a generic configuration the black
holes have different rates of energy loss and hence the mass ratios required
for mechanical equilibrium are not maintained.
This happens even in special configurations, {\it e.g.} when the temperatures are equal,
because the thermal radiation is only statistically uniform. Hence asymmetries
in the real time emission process will introduce disturbances
 driving these special configurations away from their equilibrium positions.

%%%%%%%%%%%%%%%%%%%%%%%%%%%%%%%%%%%
\subsection{Discussion of the phase structure}
\label{sec:phasestruc}
%%%%%%%%%%%%%%%%%%%%%%%%%%%%%%%%%%%

We now examine the appearance of our new multi-black hole phases in
connection to the known phases of black holes and black strings on
the cylinder (See \cite{Kol:2004ww,Harmark:2005pp,Harmark:2007md}
for reviews). In particular, as mentioned in the introduction there
is the well-known phase of the uniform black string (UBS) as well as
the non-uniform black string (NUBS), emanating from the uniform
phase at the Gregory-Laflamme point $\mu_{\rm GL}$.
Recently, numerical investigations \cite{Wiseman:2002zc,Sorkin:2004qq,Kleihaus:2006ee,%
Sorkin:2006wp,Sorkin:2003ka,Kudoh:2003ki,Kudoh:2004hs} confirmed the prediction
\cite{Kol:2002xz} that the non-uniform  phase connects
via a horizon topology changing phase transition
\cite{Kol:2002xz,Wiseman:2002ti,Kol:2003ja,Sorkin:2006wp}
to the phase of a  single localized black hole (LBH) (see Fig.~\ref{fig:phasediag}).
This point is generally referred to as the merger point.

Moreover, as reviewed in Section \ref{sec:equilibrium}, for any
solution that falls into the $SO(d-1)$-symmetric ansatz
\eqref{ansatz} of Ref.~\cite{Harmark:2002tr}, one can obtain a
copied solution \cite{Horowitz:2002dc,Harmark:2003eg,Harmark:2003yz}
by changing the periodicity of the circle from $L$ to $k L$ with $k$
an integer. As mentioned above, this includes the localized black
hole phase, from which one generates in this way the multi-black
hole solutions with $k$ equal mass black holes, which we denote by
LBH${}_k$. It also includes the non-uniform black string phase, from
which we generate copies which we denote by NUBS${}_k$, emerging
from the uniform phase at critical mass $\mu_{\rm GL}/k^{d-3}$. This
thus means that the LBH${}_k$ phase will connect to the NUBS${}_k$
phase
 via a horizon topology changing phase transition at the
$k$-copied merger point (see Fig.~\ref{fig:phasediag} for $k=2$).

We now turn to the question posed in Section \ref{sec:phasediag}:
Where do all the new multi-black hole phases end in the phase
diagram? For definiteness, let us consider again configurations with
two black holes. The LBH and NUBS phases are connected via the
topology-changing merger point, and likewise the LBH${}_2$ and
NUBS${}_2$ phases are connected via the 2-copied merger point. As
explained in Section \ref{sec:phasediag} all two-black hole
configurations with unequal mass lie (at least for small masses) in
between these two limiting phases and it is not clear where these
phases
 will end up in the phase diagram. Two scenarios where given in Section \ref{sec:phasediag},
and we now examine in more detail the possibility of the second scenario,
namely that the  black holes merge into a new non-uniform string.
Recall that this would require the temperatures of the black holes to
 approach each other at the merger point.

First of all, we have seen in Section \ref{sec:details2BH} that our
first order result for the temperatures shows that the temperatures
of the two black holes  are redshifted in such a way that they tend
to approach each other. This lends credibility to the possibility
that indeed in the full non-perturbative regime the temperatures may
converge to a common value at the merger.
If this is the case, it
seems to suggest that there would exist new non-uniform black strings
beyond the NUBS$_k$ phases, to which the unequal mass
black hole configurations could connect via new merger points.%
\footnote{Note that the original argument by Kol
\cite{Kol:2002xz,Kol:2004ww} for the merger transition of the LBH
and NUBS phases was based on Morse theory, which loosely speaking
implies that the LBH phase cannot end in ``nothing''.}

If smooth mergers do occur for different size black hole
configurations, an important question to consider is whether this a
generic feature, or if it only happens for particular
configurations. Consider for the example the case of two black
holes, for which we have two free parameters namely the total mass
and the ratio of the individual masses. Equating their temperatures
fixes the mass ratio as a function of the total mass (see
Eq.~\eqref{ratioTemp}). Similarly achieving a merger of the black
holes also fixes the ratio with another function of the total mass.
If we imagine these two functions of the mass to be independent, it
follows that we only expect these two functions to intersect at
discrete points in the space of parameters defining the
configuration. On the other hand, if these two functions are not
independent, due to the interrelation between geometry and energy in
General Relativity, one can instead imagine that the two functions
always intersect, so that the smooth mergers are a generic feature.

As discussed above, smooth mergers for different size black hole
configurations suggest that new non-uniform black string phases
exist. If this is the case, there are certain constraints on such
new phases from general arguments. Firstly, it is clearly not
possible that there are non-uniform black strings emerging from the
uniform black string in the range $\mu_{\rm GL}/2^{d-3} < \mu <
\mu_{\rm GL}$ \cite{Gregory:1993vy}. Also, it does not seem possible
that one can have other branches than the known ones coming out of
the Gregory-Laflamme point (or its $k$-copies) of the uniform black
string given the higher-order perturbative analysis of
Ref.~\cite{Gubser:2001ac}. Secondly, it is impossible to (locally)
have a continuum of non-uniform black string solutions in the phase
diagram. To prove this assertion imagine that there is a
two-dimensional continuous parameter space of solutions and consider
two points, say $A$ and $B$, in this continuum. It follows from the
continuity that one can always connect these two points by two
different paths of solutions. Imagine now that the two-dimensional
space of solutions projects into a two-dimensional region in the
$(\mu,n)$ phase diagram. If we then furthermore take the paths so
that $n$ in path 1 is greater than $n$ in path 2, then we get an
contradiction when using the Intersection Rule of
Ref.~\cite{Harmark:2003dg}. This is because  $\delta (S_1/S_2) =
(n_1-n_2) M \delta M/((d-1)T_1 T_2 S_2^2)$ where the indices on the
quantities refer to the paths. Since $n_1>n_2$, the right hand side
is strictly positive. Thus, the ratio $S_1/S_2$ in point B should be
greater than 1, but that is not possible since the two paths should
go to the same solution. We thus conclude that a locally continuous
space of solutions is impossible%
\footnote{Note that implicit in the above argument is the assumption
that there is only one connected horizon with a given temperature. Thus,
the fact that multi-black hole configurations cover a continuous region in
the phase diagram is not a contradiction because they contain disconnected horizons
typically at different temperatures.},
except in the very special case
where the continuous space of solutions projects onto a
one-dimensional subspace in the $(\mu,n)$ phase diagram. This
provides a further argument that smooth mergers would only occur at discrete
points, because there could only be a discrete set of non-uniform string solutions
to which the the merging black holes could connect.

Given these two constraints, there is still the possibility that new
non-uniform black strings may exist. Namely, it is conceivable that
the NUBS${}_k$ phases ($k \geq 1$) develop their own zero modes as
one moves away a finite distance away from the GL point (or its
$k$-copies). This is a non-perturbative effect that would not show
up in the perturbative analysis of Gubser. These zero-modes on the
non-uniform black string would in fact imply that they have some
region in which they are respectively classically unstable or
stable, just as for the uniform black string. Such a bifurcation of
new non-uniform strings from the presently known ones would also be
discrete and thereby evade the second restriction presented above.
Furthermore, in this scenario one could imagine a fractal structure
of further bifurcations into new non-uniform strings, all of which
eventually end up in a particular multi-black hole configuration. If
true, this would fit well with the smooth mergers of different size
black hole configurations occurring only at a discrete points in the
space of configurations. It would be very interesting to explore
this possibility further.

Another point that we already alluded to in Section \ref{sec:phasediag} is that we do
not expect the phases of two black hole configurations to stay in between
the LBH and LBH${}_2$ curves in the $(\mu,n)$ phase diagram (and similarly for
multi-black hole configurations with more than two black holes).
 To see this consider the LBH curve in
Fig.~\ref{fig:phasediag}. This curve has a point at which $\mu$ is maximal,
occurring well before the merger point. Beginning with this maximal mass
single-black hole configuration we can add a tiny black hole on the opposite side
of the circle and reach a two-black hole configuration with greater mass
than the original configuration. This clearly implies that
the two-black hole configurations can extend outside the wedge bounded by the
LBH and LBH${}_2$ curves. In fact, one can similarly argue by starting from the
extremal point on the LBH${}_2$ curve that by removing a tiny mass from one
of the two black holes, one can reach a two-black hole configuration to the left of
this curve. Another interesting example comes from adding a pair of tiny black holes
to any LBH${}_2$ configuration to produce a four-black hole configuration in its
neighborhood in the phase diagram, very far away from the wedge enclosed by
the LBH${}_3$ and LBH${}_4$ curves.
The above reasoning can be extended by imagining further additions of tiny
masses, in more complicated starting configurations, leading to a intricate pattern
of crossings of lines in the $(\mu,n)$ phase diagram.

We have also presented evidence in this paper for the
possibility of a new class of static
lumpy black holes in Kaluza-Klein space. Again, it would be interesting to study
this further, and examine how these in turn might connect to new non-uniform phases.

%%%%%%%%%%%%%%%%%%%%%%%%%%%%%%%%%%%
\subsection{\label{sec:fluid}A fluid analogy}
%%%%%%%%%%%%%%%%%%%%%%%%%%%%%%%%%%%

It is also interesting, though more speculative, to consider the appearance
of the multi-black hole configurations in relation to an analogue model
for the Gregory-Laflamme (GL) instability, recently proposed in Ref.~\cite{Cardoso:2006ks}.
There it was pointed out that the GL instability of a black string has a natural
analogue description in terms of the Rayleigh-Plateau (RP) instability of a fluid cylinder.
It turns out that many known properties of the gravitational  instability
have an analogous manifestation in the fluid model. These include the behavior of
threshold mode with $d$, dispersion relations, the existence of critical dimensions
and the initial stages of the time evolution%
\footnote{Recently, another feature of these instabilities has been matched.
If rotation is added to the fluid the strength of the fluid
instability increases because the centrifugal force is bigger in a
crest than in a trough of the configuration. On the gravity side it was found in
 Ref.~\cite{Kleihaus:2007dg} that rotating black strings, even for large rotation,
are indeed also unstable to the GL instability.}
(see Refs.~\cite{Cardoso:2006ks,Cardoso:2006sj,Cardoso:2007ka} for details).

Since our reasoning below relies on the time evolution of the system
and its endpoint, it is worth mentioning that the full time
evolution of the RP instability is well known (both numerically and
experimentally, see Refs.~\cite{Nayfeh,Shokoo,Stone,Eggers} for
details). On the gravity side only the initial stages of the GL
instability has been numerically studied so far
\cite{Choptuik:2003qd}. Comparing with the fluid system there is an
interesting match between the initial stage of the evolution in the
two systems. Starting from a single sinusoidal perturbation both
develop an almost cylindrical thread or neck in between the two half
rounded boundary regions. This can be confirmed by comparing Fig.~1
of \cite{Stone} (which describes the full RP evolution) and Fig.~2
of \cite{Choptuik:2003qd} (that describes the initial stage of the
GL evolution).

One should be cautious when applying the analogue model, especially
in what concerns the evolution of the systems. The reason is that
the analogy is partly based on the similarity between the first law
of black hole thermodynamics and the fluid relation $dE = T d A $
where $ E$ is the potential energy associated with surface tension
(free energy), $T$ the effective surface tension and $A$ the surface
area of the fluid. This means that both systems tend to extremize
the area. However, on the gravity side we
 know that a black object evolves such that its horizon area never decreases,
 whereas a fluid evolves toward a configuration with smaller area,
 since this decreases its potential energy.
Despite these reversed dynamical features, it is worthwhile to
notice that just like a multi-black hole system will maximize its
entropy by merging into one single black hole containing all the
mass, so will an array of fluid droplets merge into a single drop in
order to minimize its surface area at fixed volume.

Having alerted the reader to these caveats, we proceed
with the analogy in hand, considering the time evolution of the fluid
in further detail. A representative study of particular interest for our purposes was
carried out in \cite{Stone}. The main conclusion is that if we start
with a single sinusoidal perturbation in a cylindrical liquid
bridge, the higher harmonics generated by non-linear effects are
responsible for the development of a long neck that breaks%
\footnote{Note that at the pinch-off there is another similarity
that characterizes both instabilities. On the gravity side, one
would need to use quantum gravity when the pinch-off region reaches
the Planck scale and General Relativity is no longer valid. Likewise,
close to breakup of the fluid, when the radius of the liquid bridge is of molecular
size, the (continuum) hydrodynamic theory is no longer a good
approximation and simulations of the molecular dynamics  are required.}
 in a self-similar process \cite{Stone,Nayfeh,Shokoo}. We end up with an
array of satellite drops with different sizes. Hence, if  the correspondence
indeed extends to the full evolution, the multi-black holes would
be the natural gravity analogues of the main drop and satellite
droplets array observed in the fluid analysis.

 Furthermore
the analogue model would thus argue in favor of the scenario in which
 the neutral black string will pinch off.
 Moreover, the multi-black hole configurations constructed in this paper
 would play an important role in the intermediate stages of the GL instability.
It would be interesting to examine this application of the analogue fluid model
and its consequences more closely.

%%%%%%%%%%%%%%%%%%%%%%%%%%%%%%%%%%%
\subsection{Outlook}
\label{sec:outlook}
%%%%%%%%%%%%%%%%%%%%%%%%%%%%%%%%%%%

The study of Kaluza-Klein black holes and their high degree of non-uniqueness can
be viewed in the broader context of studying black objects in higher dimensional gravity.
Here, research in the last years has revealed that also in asymptotically flat space
 a very rich phase structure of stationary black objects is expected. In particular,
in five-dimensional Einstein gravity there exists, beyond the
rotating Myers-Perry black hole, a black ring solution
\cite{Emparan:2001wn} (see \cite{Emparan:2006mm} for a review).
Recently further new stationary solutions, called `black Saturns'
\cite{Elvang:2007rd,Iguchi:2007is}, have been constructed explicitly
in five-dimensional gravity. These solutions, consisting of a
spherical black hole with black rings around it, are similar to the
multi-black hole configurations, in that the generic solution is not
in thermal equilibrium, with different temperatures for each
connected component of the event horizon. Furthermore, one may
compare the configurations with highest entropy in the two systems.
It was shown in Ref.~\cite{Elvang:2007hg} that the maximal entropy
configuration for fixed mass and angular momentum consists of a
central, close to static, black hole and a very thin black ring
around it. For any value of the angular momentum, the upper bound on
the entropy is then equal to the entropy of a static black hole of
the same total mass. These maximal entropy black Saturns are not in
thermal equilibrium. In some sense the same features are observed
for multi-black hole configurations. If we restrict to the case of
two black holes, the highest entropy configuration (see Section
\ref{sec:details2BH}) is that of an infinitesimally small black hole
together with a large black hole, \ie far away from thermal
equilibrium. The entropy of that configuration is bounded from above
by that of a single black hole of the same mass.

It is also worth emphasizing that the solution technique employed in
this paper can be applied to other black hole systems where one
lacks the symmetries or other insights to construct exact solutions.
The general idea is to identify a suitable perturbation parameter of
the putative solution, and follow similar steps as outlined in
Section \ref{sec:construction}.

Another  open direction to pursue is to apply numerical techniques to
extend the construction of multi-black hole configurations into the non-perturbative
regime, as was successfully done for a single black hole on a cylinder in five and six dimensions
\cite{Sorkin:2003ka,Kudoh:2003ki,Kudoh:2004hs}.
Such an analysis could confirm whether indeed there are multi-black
hole configurations for which the temperatures converge when approaching
the merger points as one increases the mass, as was
discussed in Section \ref{sec:details2BH}. Furthermore, it
is possible that in this way one could confirm the existence of
the lumpy black holes conjectured in Section \ref{sec:details3BH}, where
we recall that these are most likely for higher dimensions.

A further, but technically complicated, direction to pursue is to
extend the solutions of this paper to the next, \ie second, order.
For the case of a single black hole in five dimensions, the second
order correction to the metric and thermodynamics have been studied
in \cite{Karasik:2004ds}. More generally, the second order
correction to the thermodynamics was obtained in
Ref.~\cite{Chu:2006ce} for all $d$ using an effective field theory
formalism in which the structure of the black hole is encoded in the
coefficients of operators in an effective worldline Lagrangian. It
would be interesting to obtain the second-order corrected metric and
thermodynamics for the multi-black hole case considered in this
paper.

There are also potential applications related to string theory and gauge
theory.
It is known that the phases of Kaluza-Klein
black holes are related via a boost/U-duality map \cite{Harmark:2004ws}
(see also \cite{Bostock:2004mg,Aharony:2004ig})
  to phases of non- and near-extremal branes on a transverse
circle, appearing as solutions in type II string theory or M-theory.
Via the gauge/gravity correspondence
\cite{Maldacena:1997re,Aharony:1999ti} this has implications for the
phase structure of the dual non-gravitational theories at finite
temperature. For instance, it is possible to obtain in this way
non-trivial predictions
\cite{Aharony:2004ig,Harmark:2004ws,Harmark:2005dt} about the strong
coupling dynamics of supersymmetric Yang-Mills theories on compact
spaces and of the thermal behavior of little string theory.

  As an important example, Ref.~\cite{Aharony:2004ig}
considered finite temperature two-dimensional supersymmetric
Yang-Mills on a spatial circle, which by the boost/U-duality map is
related to the phase structure of Kaluza-Klein black holes in ten
dimensions. The corresponding phase structure that is present at
strong coupling in the two-dimensional Yang-Mills theory on the
torus $S^1_\beta \times S^1$ was then qualitatively matched to the
phase structure in the weakly coupled gauge theory. In particular,
it was found in \cite{Aharony:2004ig} that the eigenvalue
distribution of the spatial Wilson loop distinguishes between the
three different phases seen at strong coupling: The uniform phase
corresponds to a uniform eigenvalue distribution, the non-uniform
phase corresponds to a non-uniform eigenvalue distribution and the
localized phase maps to a gapped eigenvalue distribution. It would
be interesting to see if there are also multiply gapped eigenvalue
distributions (see e.g. Ref.~\cite{Jurkiewicz:1982iz}), corresponding to the localized phase of multi-black
holes found in this paper. While those would probably be unstable as
mentioned above, they may still appear as unstable saddle points.

Finally, we remark on an open direction that is related to
microscopic calculations of the entropy of black holes. In
Ref.~\cite{Harmark:2006df} (see \cite{Harmark:2007uy} for a short
summary) the boost/U-duality map of \cite{Harmark:2004ws} was
extended to the case of branes with more than one charge. One of the
results is that by starting with neutral Kaluza-Klein black holes in
five dimensions  one can generate five-dimensional three-charge
black holes on a circle, obtained from corresponding three-charge
brane configurations in type II/M-theory via compactification. In
particular, when one applies this map to a single neutral localized
black hole one obtains a three-charge black hole localized on the
transverse circle. For this case, it was shown that in a partial
extremal limit with two charges sent to infinity and one finite, the
first correction
to the finite entropy is in agreement with the microscopic entropy.%
\footnote{The entropy matching for the single three-charge black hole
case considered in \cite{Harmark:2006df}  was extended in
Ref.~\cite{Chowdhury:2006qn} to second order.}
By applying the map to the multi-black hole solutions of this paper one will
generate three-charge multi-black holes on a circle. The results
of Section \ref{sec:thermodynamics} can then be used to compute the first correction to
the finite entropy of these three-charge multi-black hole configurations,
and it would be interesting to then derive these expressions from
a microscopic calculation as well.
Furthermore, in Ref.~\cite{Chowdhury:2006qn}
a simple microscopic model was proposed that reproduces most of the
features of the phase diagram of three-charge black holes on a circle,
including the new non-uniform phase. It would be interesting to see
if this model can also account for the corresponding localized
three-charge multi-black hole solutions.

%%%%%%%%%%%%%%%%%%%%%%%%%%%%%%%%%%%%%%%%%%%%%%%%%%%%%%%%%%%%
\section*{Acknowledgements}

We thank Henriette Elvang, Roberto Emparan, Gary Horowitz, Veronika Hubeny, Barak Kol,
Don Marolf, Mukund Rangamani, Evgeny Sorkin and Toby Wiseman for useful
discussions. The authors would like to thank the KITP for
hospitality during the program ``Scanning new horizons: GR beyond 4
dimensions'', where this work was started. The authors would also
like to thank the organizers of the workshops
 "Einstein's Gravity in Higher Dimensions" Jerusalem Feb. 18-22, 2007,
 and "Pre-strings 2007" Granada June 18-22, 2007  where part of this work was done.
 OD acknowledges the hospitality of the Perimeter Institute during early stages
 of this work.
This work was partially funded by Funda\c c\~ao para a Ci\^encia e
Tecnologia (FCT, Portugal) through project PTDC/FIS/64175/2006. OD
acknowledges financial support provided by the European Community
through the Intra-European Marie Curie contract MEIF-CT-2006-038924.
The work of TH and NO is partially supported by the European
Community's Human Potential Programme under contract
MRTN-CT-2004-005104 `Constituents, fundamental forces and symmetries
of the universe'. TH would like to thank the Carlsberg Foundation
for support. Research at the Perimeter Institute  is supported in
part by the Government of Canada through NSERC and by the Province
of Ontario through MRI. RCM also acknowledges funding from an NSERC
Discovery grant and the Canadian Institute for Advanced Research.
Research at the KITP was supported in part by the NSF under Grant
No. PHY05-51164. % v2

%%%%%%%%%%%%%%%%%%%%%%%%%%%%%%%%%%%%%%%%%%%%%%%%%%%%%%%%%%%
\begin{appendix}

%%%%%%%%%%%%%%%%%%%%%%%%%%%%%%%%%%%%%%%%%%%%%%%%
\section{\label{sec:2unequal}Data for two unequal mass black holes}
%%%%%%%%%%%%%%%%%%%%%%%%%%%%%%%%%%%%%%%%%%%%%%%%

In this appendix we give some useful explicit expressions for the
quantities that are involved in the construction of the two-black
hole configuration, further discussed in Section
\ref{sec:details2BH}.

The mass fractions of the two black holes are taken as $\nu_1 =
\frac{1}{2} + \kappa $, $\nu_2 = \frac{1}{2} - \kappa $ and the
equilibrium configuration is chosen such that the first black hole at $z_1^*=0$ and
the second at $z_2^*=\pi$, \ie at opposite points on the circle. The
function \eqref{def F}  entering the Newtonian potential is then
given by
\begin{equation}
 F (r,z) = 2^{d-3} {\cal{F}} (2r,2z) + 2^{d-2} \kappa \  \hat{\cal{F}} (2r,2z) \,,
\end{equation}
where we have defined
\begin{equation}
\label{Ffun} {\cal{F}} (r,z) \equiv \sum_{m=-\infty}^\infty \frac{ 1
}{ [r^2 + (z - 2\pi m)^2]^{\frac{d-2}{2}}} \spa \hat{\cal{F}} (r,z)
\equiv \sum_{m=-\infty}^\infty \frac{ (-1)^m }{ [r^2 + (z - 2\pi
m)^2]^{\frac{d-2}{2}}}\,.
\end{equation}
The function ${\cal{F}} (r,z)$ is identical to the one entering the
Newtonian potential for the single black hole case, and details can
be found in Appendix B of \cite{Harmark:2002tr}. Using Poisson
resummation the large $r$ expansions of the two functions in
\eqref{Ffun} are obtained as
\begin{eqnarray}
{\cal{F}} (r,z)=\frac{k_d}{r^{d-3}} {\biggl (} 1+2
 \sum_{m=1}^{\infty} h(m r)\cos (m z ) {\biggr )} \,,
 \label{Fcal}
\end{eqnarray}
\begin{equation}
\label{Fhat} \hat{\cal{F}} (r,z) = \frac{2 k_d}{r^{d-3}}
\sum_{m=0}^\infty h( [m+1/2]  r)\cos( [ m+1/2] z )  \,,
\end{equation}
where $k_d$ and $h (x)$ are defined in \eqref{defkd}, \eqref{defh} respectively.

Note that for even $d$, the relevant Bessel function takes the form of a
polynomial of finite degree.
This allows to write explicit expressions for ${\cal F}(r,z)$ and
$\hat {\cal F} (r,z)$ (and similarly for $v(r,z)$ in \eqref{fourier
v}). For example, for $d=4$ one has
\begin{eqnarray}
 {\cal{F}} (r,z) = \frac{1}{2r} \frac{ \sinh r}{\cosh r - \cos
z}\,, \qquad
  \hat{\cal{F}} (r,z) = \frac{1}{r} \frac{\sinh (r/2) \cos (z/2)}{\cosh r -
\cos z}\,, \qquad {\rm for}\:\: d=4\,.
\end{eqnarray}
For $d=6$ one has
\begin{eqnarray}
{\cal{F}} (r,z) &=& \frac{1}{4r^3} \frac{ \sinh r}{\cosh r -\cos z}
+\frac{1}{2r^2}
\frac{\sinh^2(r/2)\cos^2(z/2)-\cosh^2(r/2)\sin^2(z/2)}
 {(\cosh r-\cos z)^2}\,,
\nonumber \\
\hat{\cal{F}} (r,z) &=& \frac{1}{2 r^3} \frac{\sinh (r/2) \cos
(z/2)}{\cosh r - \cos z} +\frac{1}{8r^2}
\frac{\sinh^2(r/4)\cos^2(z/4)-\cosh^2(r/4)\sin^2(z/4)}
 {[\cosh (r/2)-\cos (z/2)]^2} \nonumber \\
 & &+ \frac{1}{8r^2}
\frac{\cosh^2(r/4)\cos^2(z/4)-\sinh^2(r/4)\sin^2(z/4)}
 {[\cosh (r/2)+\cos (z/2)]^2} \,, \qquad {\rm for}\:\: d=6\,.
\end{eqnarray}

For the small $\rho$ expansion, we first present the results for
general location $z_2^*$ of the second black hole, restricting to
the equilibrium configuration $z_2^*= \pi $ at the end. In the
region near the first or second black hole respectively, we have
from \eqref{lim F} the expansions
\begin{equation}
F(r,z) \simeq \frac{ \frac{1}{2} + \kappa }{\rho^{d-2}} +
\Lambda^{(1)} \spa\qquad F(r,z) \simeq \frac{ \frac{1}{2} - \kappa
}{\rho^{d-2}} + \Lambda^{(2)}\,,
\end{equation}
where $\Lambda^{(1,2)}$ are computed from \eqref{lambda}
\begin{eqnarray}
& & \Lambda^{(1)} =    \frac{(\frac{1}{2} + \kappa) 2\zeta (d-2)}{
(2 \pi)^{d-2}}
   +  \frac{\frac{1}{2} - \kappa}{(z_2^*)^{d-2}}+\frac{\frac{1}{2} - \kappa} { (2 \pi)^{d-2}}
  \left[ \zeta \left(d-2,1+\frac{z_2^*}{2\pi}\right)
  +\zeta \left(d-2,1-\frac{z_2^*}{2\pi}\right) \right]
   \,, \nonumber \\
& & \Lambda^{(2)} =  \frac{(\frac{1}{2} - \kappa)2\zeta (d-2)}{ (2
\pi)^{d-2}}
  + \frac{\frac{1}{2} + \kappa}{(2\pi-z_2^*)^{d-2}}+\frac{\frac{1}{2} + \kappa} { (2 \pi)^{d-2}}
  \left [ \zeta \left(d-2,2-\frac{z_2^*}{2\pi}\right)
  +\zeta \left(d-2,\frac{z_2^*}{2\pi}\right) \right
  ]  \,, \nonumber \\
  \label{lambda2BHs}
\end{eqnarray}
and we recall the definitions \eqref{Gzeta}, \eqref{defZi}. In
particular, for the equilibrium configuration $z_2^* = \pi$ these
expressions reduce to
\begin{equation}
\label{lambda2BHs0} \Lambda^{(1)} = [ 2^{d-3} + 2 \kappa
(1-2^{d-3})] \frac{2 \zeta(d-2)}{(2\pi)^{d-2}} \spa\quad
\Lambda^{(2)} = [ 2^{d-3} - 2 \kappa (1-2^{d-3})] \frac{2
\zeta(d-2)}{(2\pi)^{d-2}}\, .
\end{equation}
Note that for $\kappa =1/2$, the expression for  $\Lambda^{(1)}$
reduces to the correct result for a single black hole. Finally, we
record the sum
\begin{eqnarray}
\sum_{i=1}^{2} \nu_i \Lambda^{(i)}=2^{d-3}\, \frac{2\zeta (d-2)}{(2
\pi)^{d-2}} \left[ 1+2^{5-d}\left( 1-2^{d-3}\right)\kappa^2\right],
  \label{sum2bh}
\end{eqnarray}
which is used in the text to compute various thermodynamic
quantities.

%%%%%%%%%%%%%%
\end{appendix}

%\bibliographystyle{utphys}
%\bibliography{bibrot}
%\end{document}

\providecommand{\href}[2]{#2}\begingroup\raggedright\endgroup

\end{document}